\documentclass[sigconf]{acmart}

\AtBeginDocument{%
  \providecommand\BibTeX{{%
    \normalfont B\kern-0.5em{\scshape i\kern-0.25em b}\kern-0.8em\TeX}}}

\usepackage[english]{babel}
\usepackage{blindtext}
\usepackage{siunitx}
\usepackage{graphicx}
\usepackage{subcaption}
\usepackage{balance}

\usepackage{algorithm}
\usepackage[noend]{algpseudocode}

\usepackage{booktabs}
\usepackage{multirow}
\usepackage{url}
\usepackage{tablefootnote}
\usepackage{varwidth}

\setcopyright{none}
\renewcommand\footnotetextcopyrightpermission[1]{}

\newcommand{\model}{\textsf{VAD360}}

\begin{document}

\title{\model{}: Viewport Aware Dynamic 360-Degree Video Frame Tiling}

\author{Chamara Kattadige}
\email{ckat9988@uni.sydney.edu.au}
\affiliation{%
\institution{The University of Sydney}
\city{Sidney}
\country{Australia}
}

\author{Kanchana Thilakarathna}
\email{kanchana.thilakarathna@sydney.edu.au}
\affiliation{%
\institution{The University of Sydney}
\city{Sydney}
\country{Australia}
}


\begin{abstract}
  \ang{360} videos a.k.a. spherical videos are getting popular among users nevertheless, omnidirectional view of these videos demands high bandwidth and processing power at the end devices. Recently proposed viewport aware streaming mechanisms can reduce the amount of data transmitted by streaming a limited portion of the frame covering the current user viewport (VP). However, they still suffer from sending a high amount of redundant data, as the fixed tile mechanisms can not provide a finer granularity to the user VP. Though, making the tiles smaller can provide a finer granularity for user viewport, high encoding overhead incurred. To overcome this trade-off, in this paper, we present a computational geometric approach based adaptive tiling mechanism named \model{}, which takes visual attention information on the \ang{360} video frame as the input and provide a suitable non-overlapping variable size tile cover on the frame. Experimental results shows that \model{}  can save up to 31.1\% of pixel redundancy before compression and 35.4\% of bandwidth saving compared to recently proposed fixed tile configurations, providing tile schemes within 0.98($\pm 0.11$)s time frame.
\end{abstract}

\ccsdesc[500]{Human-centered computing~Virtual reality}


\keywords{\ang{360}/VR video streaming, Video frame tiling}

\maketitle

\vspace{-2mm}
\section{Introduction}

Over the past decade, video streaming has been dominating the global internet traffic accounting for 80\% of total traffic~\cite{schmitt2019inferring}. Among them, \ang{360} vidoes, a.k.a. spherical videos, is becoming increasingly popular as it enables immersive streaming experience. Along with the advancement of Head Mount Devices (HMDs) (e.g., Facebook Oculus~\cite{oculus}, Microsoft Hololens~\cite{hololens}) and smartphones, content providers (e.g., YouTube (YT)~\cite{yt} and Facebook (FB)~\cite{fb}) have already facilitated both on-demand and live-streaming of \ang{360} videos.

Despite the gaining popularity of \ang{360} videos, its intrinsic larger spherical view has posed several challenges against providing expected high user Quality of Experience (QoE). Firstly, \ang{360} videos demand high network bandwidth since spherical video frames should be 4--6 times larger than normal video frames to achieve the same user perceived quality~\cite{qian2018flare}. Secondly, processing such video frames at end devices imposes high resource utilization is problematic particularly with the limited resources such as in smartphones. Finally, \ang{360} video streaming requires strictly low latency, which should be within the \textit{motion-to-photon delay}\footnote{Delay between user head movement and the finishing rendering the corresponding VR frame on to the display} (just above 25ms), otherwise the user may suffer from severe motion or cyber-sickness~\cite{liu2018cutting}.  

Recently proposed viewport (VP)
--current visbile area, typically encompass $\ang{100}\times \ang{100}$ Field of View (FoV)--
adaptive streaming~\cite{xiao2017optile}, which streams only a selected portion of the \ang{360} video frame, has shown a great promise in addressing the above issues. Especially, the tiling mechanism, which divides the entire frame into tiles and send only the tiles within the user VP can reduce the  amount of data transferred to the client and increases the user perceived video quality compared to streaming entire video frame~\cite{qian2018flare,he2018rubiks,xie2017360probdash,xie2017poi360}. However, due to the fix number of tiles (typically ranges between (24--36))~\cite{qian2018flare,he2018rubiks} and their fixed sizes, optimum gain of bandwidth saving is not achieved by these mechanisms. On the one hand, fixed tiling schemes fails to provide finer boundary to the user FoV, therefore, there is a high pixel redundancy~\cite{xiao2017optile}. On the other hand, they do not have awareness of visually attractive regions and FoV distortion on Equirectangular Projection (ERP) when encoding the tiles. 
For example, 
corresponding regions of the polar regions of the spherical view on the ERP frame are highly distorted and have less viewing probability. However, fixed tiling encodes these regions at the same bit-rate levels and in smaller tile size as in the equatorial regions, adding unnecessary quality overhead and losing many compression opportunities compared to having bigger tiles~\cite{xiao2017optile,hooft2019tile}

To this end, provisioning an viewport aware adaptive tiling mechanism with variable sized tiles can provide fine granularity for the FoV and also increase compression gain. This enables content providers to reduce encoding overhead incur by tiling the videos and identify the tiles which should be in high quality at a fine granularity. At the network provider side, transmitted data volume will be further decreased reducing the bandwidth consumption for \ang{360} video streaming whilst providing opportunities to re-innovate caching mechanisms for \ang{360} videos. Finally, existing DASH (Dynamic Adaptive Streaming over HTTP) protocols on \ang{360} videos can be enhanced to provide better user QoE at the client side.

Achieving such an adaptive tiling mechanism is challenging due to sheer volume of videos need to be processed, stored and streamed from the content providers perspective. 
In on aspect, running algorithms to generate suitable tile schemes should not demand high processing power, as the servers are already in high utilization to support excessive demand of current video streaming services. In another aspect, algorithms itself should process in a minimal time period. However, recently proposed dynamic tiling solutions~\cite{xiao2017optile,zhou2018clustile,ozcinar2019visual} compromise both these aspects by encoding all possible tile schemes for a given frame even in more than the quality levels provided in DASH protocols~\cite{ozcinar2019visual}. Furthermore, Integer Liner Programming (ILP) based solutions and exhaustive searching mechanisms on all possible solutions provided in these proposals requires a significant processing time. 

In this paper, we propose \model{} (\textbf{V}iewport \textbf{A}daptive \textbf{S}calable 360-degree Video Frame \textbf{Til(e)}ing), an adaptive tiling mechanism for \ang{360} video frames supported by dynamic nature of user VPs in \ang{360} video streaming. In \model{}, we leverage a computational geometric approach, which we call \textit{Minimal Non-overlapping Cover (MNC)} algorithm~\cite{ohtuski1982,sun2017weighted}, to devise a suitable tile scheme by partitioning rectilinear polygons generated by combining basic tiles from $10\times 20$ grid overlaid on the \ang{360} video frame. To generate the rectilinear polygons, \model{} consists with semi-automated thresholding mechanism, which divides the \ang{360} video frame into multiple sub regions based on the visual attraction level, i.e. saliency, of pixels of the frame. 
Moreover, taking FoV distortions on the ERP frame and removing potential overlaps, \model{} further reduces the downloaded data volume, transmitted pixel redundancy and processing time for end-to-end tile generation process. 

We leverage 30, \ang{360} videos in 1 min duration with 30 user VP traces in each to build \model{} and validate its performance. Our experimental results shows that \model{} can save up to 31.1\% of pixel redundancy before compression and 35.4\% of bandwidth saving compared to recently proposed fixed tile configurations (cfgs.). Moreover, circumventing the time consuming process of encoding all possible tile combinations for exhaustive/ILP based searching algorithms, \model{} is able to generate suitable tile scheme with in avg. of 0.98 ($\pm 0.11$)s processing time with at least 80\% of individual user VP coverage in high quality. Artficats of the work are--\textit{hidden for double blind submission}--.

\section{Related work}

\textbf{VP-aware \ang{360} video streaming:}
Plethora of works have been done in VP-aware \ang{360} video streaming optimization~\cite{qian2018flare,he2018rubiks,xie2017360probdash,xie2017poi360,bao2016shooting,shi2019freedom}. In these mechanisms, a predicted user VP is sent to the content servers and a selected portion of the frame covering the requested VP is transmitted to the client. The most prominent way of selecting the required portion is
first, dividing the entire frame into fixed number of tiles and select only the tiles fall with in the user VP~\cite{qian2018flare,he2018rubiks,xie2017360probdash,xie2017poi360}. 
The overarching goal of VP aware streaming is to reduce the amount of data transmitted through the network and increase the quality of the streamed portion to increase the user perceived QoE. 
A major drawback of these proposals is that tiles in fixed size transmit high amount of redundant data as they can not provide finer boundary to the user FoV. Although,
smaller size tiles can create a finer boundary, it increases the encoding overhead which results in higher bandwidth consumption~\cite{xiao2017optile,hooft2019tile}. 


\noindent
\textbf{Adaptive tiling schemes on \ang{360} video frames:}
In contrast to the uniform size tiling, variable size tile schemes are proposed in~\cite{yu2015content,li2016novel}. They divide the frame into fixed num. of tiles but vary their size according to the latitude. Diversely, combining set of basic tiles in fixed minimum size to form larger tiles in variable amount and size are presented in~\cite{xiao2017optile, zhou2018clustile,ozcinar2019visual}. Both~\cite{xiao2017optile}~and~\cite{zhou2018clustile} leverage ILP based solution to find the best tile cfgs. taking the server side storage size and streaming data volume 
as the cost functions. Ozcinar \textit{et al.} present a exhaustive searching method 
to derive tiles while allocating a given bitrate budget~\cite{ozcinar2019visual}.
These approaches are lack of scalability due to two reasons. 
First, they require to encode all possible tile schemes, which may exceed 30000 of solutions~\cite{xiao2017optile} incurring high encoding time. Second, algorithms such as exhaustive searching/ILP  itself need longer processing time. 

\emph{In contrast to the above method, we propose a scalable, adaptive tiling mechanism leveraging a computational geometric approach, which can provide high quality tiles in user VP, whilst reducing the pixel redundancy before compression and streamed volume of data.}


\section{Background and Motivation}

\begin{figure}[h]
  \centering
  \captionsetup{}
  \begin{subfigure}{.48\columnwidth}
    \centering
    \includegraphics[trim=0 0.5cm 0 0 ,clip,width=\linewidth]{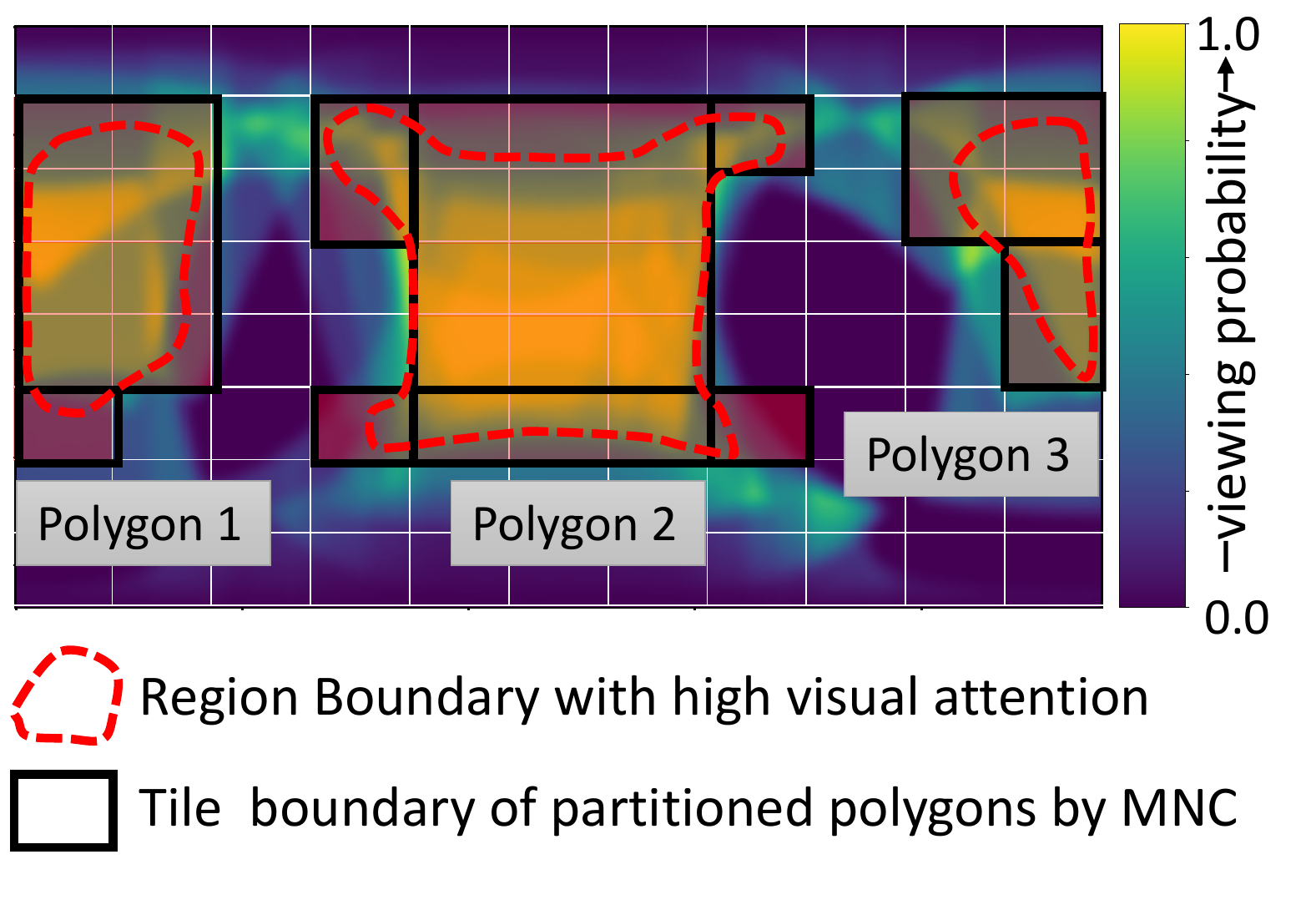}\vspace{-1mm}
    
    \caption{Partitioning regions with high visual attention using basic \textit{MNC} algorithm}
    \label{fig:mnc on vp}
  \end{subfigure}
  \hfill
  \begin{subfigure}{.48\columnwidth}
    \centering
    \includegraphics[trim=0 0.5cm 0 0 ,clip,width=\linewidth]{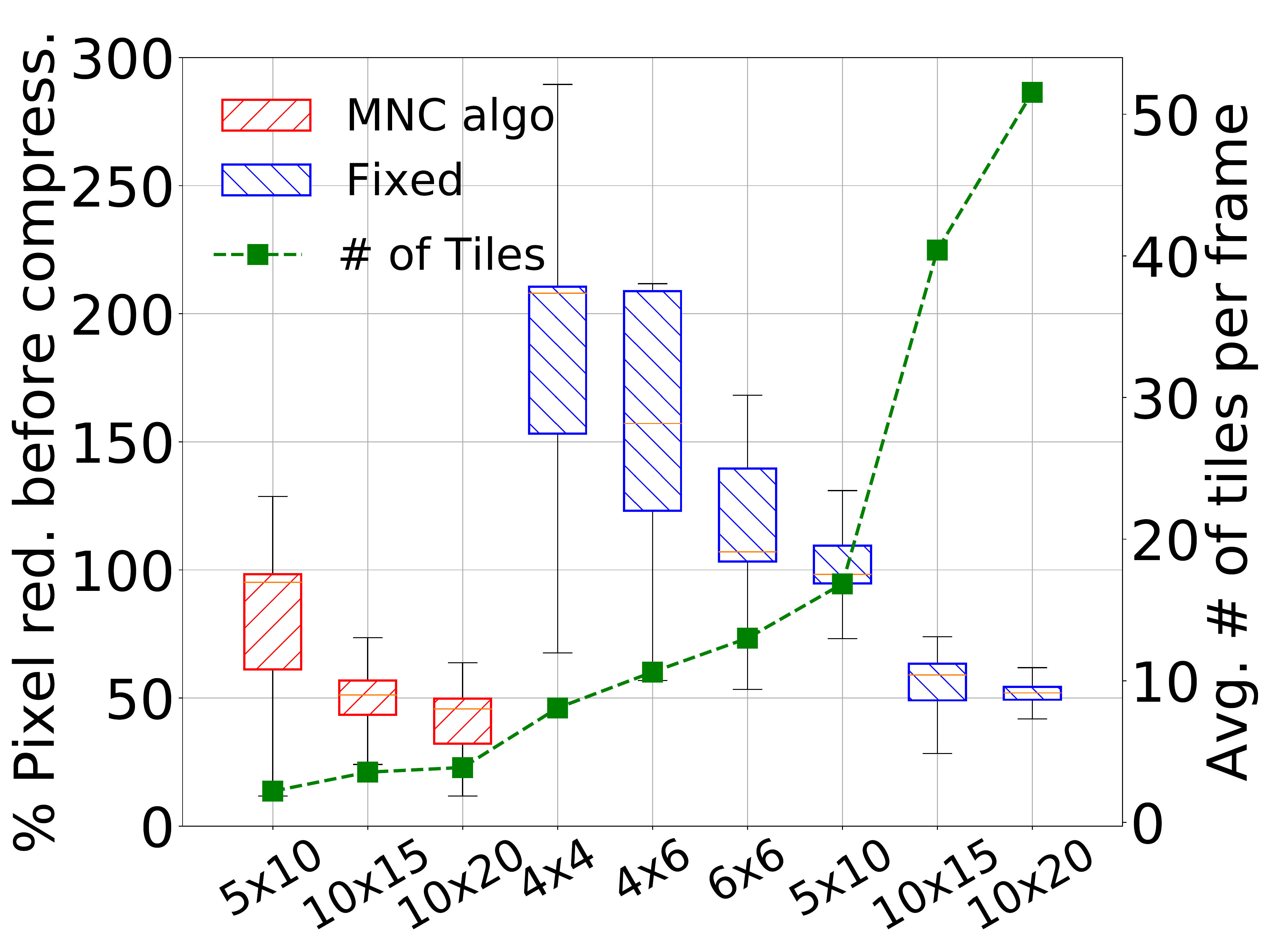}\vspace{-3mm}
    
    \caption{Pixel red. before compression \& no. of tiles per frame}
    \label{fig:comparison fixed tiles}
  \end{subfigure}
  
\vspace{-4mm}
\caption{\textit{MNC} based partitioning and Comparison with fixed tile cfgs. }
\label{fig:mnc validation}
\end{figure}

The overarching goal in \model{} is to partition video frames leveraging visual attention level of each pixel in the frame as shown in Fig.~\ref{fig:mnc on vp}. Generating visual attention maps has been well studied in the literature~\cite{nguyen2018your,monroy2018salnet360,ozcinar2019visual,xiao2017optile}, which are often developed by either analysing content features~\cite{nguyen2018your,monroy2018salnet360} or/and by analysing past VP traces of the video~\cite{ozcinar2019visual,xiao2017optile,nguyen2018your}. The primary challenge addressed in this work is how we find optimal partitioning of frames given a visual attention map. For this purpose, \model{}  leverages computational geometric approach proposed in \cite{ohtuski1982,wu1994fast}.

We now presents basics of the approach and its applicability for \ang{360} video frame partitioning. Moreover, our preliminary experiments on individual user viewports shows how efficient this utilized algorithm in frame partitioning compared to the fixed tile configurations in terms of \% pixel redundancy before compression.

\subsubsection{MNC algorithm}\label{subsubsec:basic mnc}
The original rectilinear polygon partitioning algorithm proposed in \cite{ohtuski1982,wu1994fast} creates a rectangular tile cfg. covering the polygon region with fewest number of tiles in variable size, a.k.a. \textit{Minimal Overlapping Cover (MNC)}. In breif basic \textit{MNC} algorithm runs on hole free rectilinear polygons first taking the concave (and convex) vertices in polygon boundary to find maximum independent chords. These independent chords divide the polygon into sub-rectilinear polygons which can be further partitioned by drawing chords (vertically/horizontally) from the concave vertices, from which a chord has not been drawn yet.
We refer interesting users to 
\cite{ohtuski1982,wu1994fast} for detail implementation of \textit{MNC} algorithm.

\vspace{-3mm}

\vspace{-6mm}
\subsubsection{Applicability of MNC algorithm for partitioning}\label{subsec:basic_mnc}
Fig.~\ref{fig:mnc on vp} shows that on the visual attention maps generated using past VP traces, we can create rectilinear polygons surrounding the regions with high VP probability. These polygons, which are comprised of set of basic tiles (i.e., smallest tile that can not be be further partitioned denoted as BT) can be partitioned by the  \textit{MNC} algorithm as in Section.~\ref{subsubsec:basic mnc}.
However, \textit{MNC} algorithm is unaware of visual attention as information of vertices
of polygon is enough for the partitioning. Leveraging simple pre/post-processing steps on the video frame, we can inject viewport awareness to the  \textit{MNC} algorithm converting the process for quality adaptive tiling.
which is further elaborated in Section~\ref{subsec:model overview}.




\begin{figure*}[t!]
    \centering
    \includegraphics[trim=0 0.2cm 0 0 ,clip,width=0.9\textwidth]{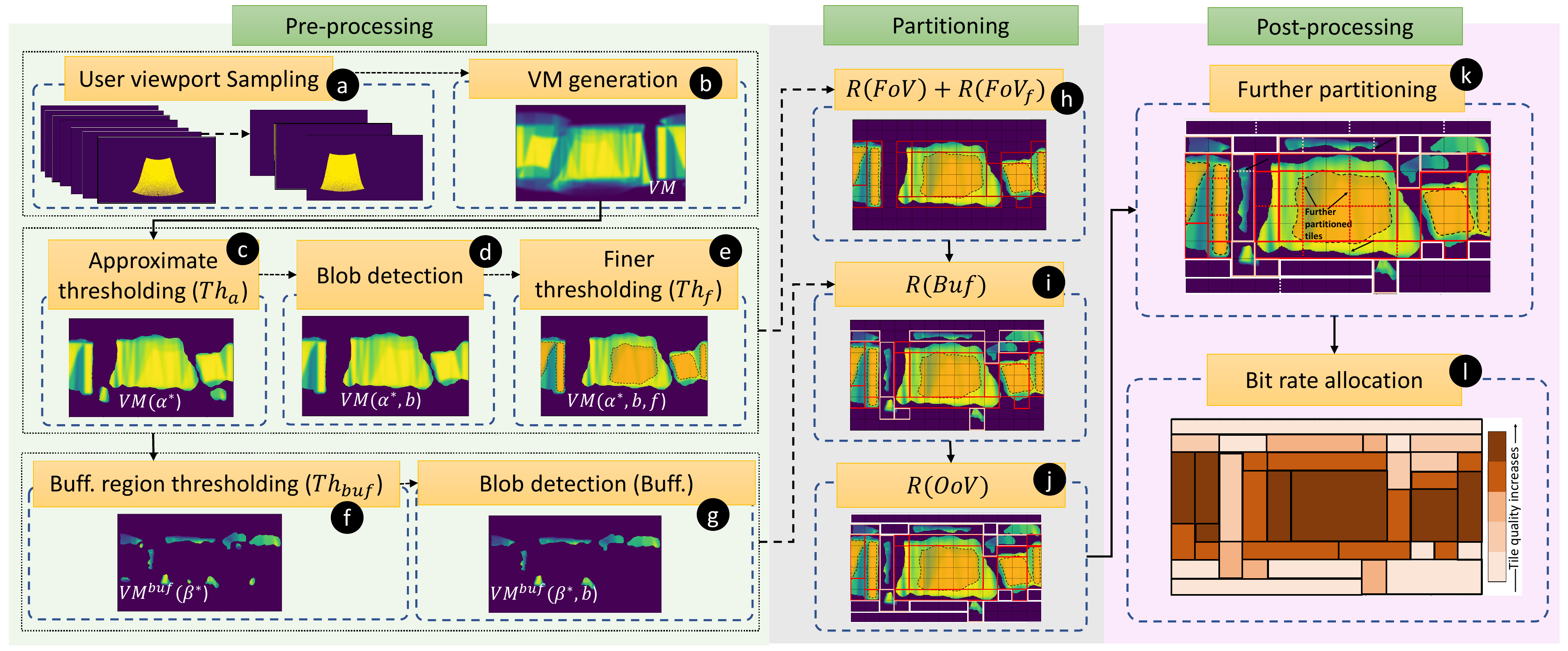}\vspace{-3mm}
    \caption{Overall architecture of \model{} including Pre-processing, Partitioning and Post-processing steps.}\vspace{-3mm}
    \label{fig:main_block}
\end{figure*}

\begin{figure}[t]
  \centering
    \centering
    \includegraphics[trim=0 0.2cm 0 0 ,clip,width=\linewidth]{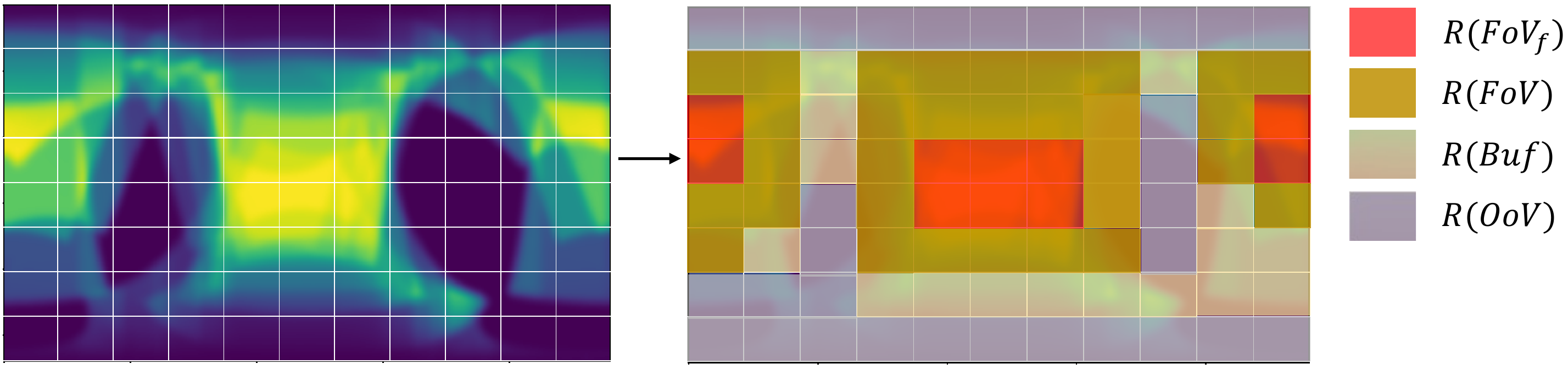}
    \vspace{-3mm}
    \caption{Four FoV regions considered in frame partitioning.}\vspace{-4mm}
    \label{fig:four_regions} 
\end{figure}

\vspace{-1mm}
\subsubsection{Comparison with fixed tile configuration}\label{subsec:basic_mnc}
We run \textit{MNC} algorithm on individual user viewports\footnote{Unlike the visual attention maps, which combines multiple user VPs, an individual user VP is a binary map representing FoV by 1 and outside the FoV by 0.} and compare the generated tile schemes with fixed tile cfgs.~\cite{qian2018flare,he2018rubiks,lo2017360} measuring the \% pixel redundancy before compression. We leverage VP maps of randomly selected 5 users from 5 sample videos representing the categorization provided in~\cite{carlsson2020had}\footnote{\cite{carlsson2020had} put videos in into 5 categories: riding, moving focus, exploration, rides and miscellaneous (a combination of previous types)}.  
We consider three basic tile configurations ($5\times 10$, $10\times 15$ and $10\times 20$) to generate rectilinear polygons as the input to the \textit{MNC} algorithm.
We compare the derived tile schemes from \textit{MNC} algorithm with five fixed tile cfgs. ($4\times 6$, $6\times 6$, $5\times 10$, $10\times 15$ and $10\times 20$) applied on the same user viewport. We measure \% pixel redundancy before compression as in Eq.~\ref{eq:pixel redundancy}. 

\vspace{-2mm}
\begin{equation}\label{eq:pixel redundancy}
    Pixel\ red.\ before\ compression = \frac{N_T - N_{FoV}}{N_{FoV}}\%
\end{equation}
\noindent
where $N_T$ and $N_{FoV}$ represent number of non-zero pixels in tiles which overlaps with the user FoV and considered FoV region respectively.
Fig.~\ref{fig:comparison fixed tiles} shows the analysis results. We see that 
$10\times 20$ cfg. results in lowest \% pixel redundancy for both \textit{MNC} partitioning and fixed tile approach. However, to cover the same single user VP, $10\times 20$ cfg. in fixed tiling requires $\times25$ the no. of tiles generated by \textit{MNC} algorithm. Therefore, compared to \textit{MNC} algorithm
, more encoding overhead incur on fixed tile cfgs. as the tile sizes are smaller and higher in amount~\cite{xiao2017optile,hooft2019tile}.




\vspace{-1mm}
\section{Design of \model{} Framework}



Now, we present \model{} holistic architecture, which includes \textit{Pre-processing}, \textit{Partitioning} and \textit{Post-processing} of \ang{360} video frames as illustrated in Fig.~\ref{fig:main_block}.


\vspace{-2mm}
\subsection{\model{} overview}\label{subsec:model overview}

The objective of the \textit{Pre-processing} step is to identify different regions in a frame, according to the expected visual attention in order to inject viewport awareness to the \textit{MNC} algorithm and reduce the complexity of detected polygons for partitioning. We first create averaged \textit{Viewport Map (VM)} combining individual user viewport maps and then apply a hierarchical thresholding mechanism to detect visual attention blobs\footnote{Regions with (near)concentric user VPs} for the following four regions, as depicted in Fig.~\ref{fig:four_regions}. The first two regions are defined on the area which covers at least 80\% of the user VPs namely, $R(FoV_f)$ which covers the most attractive regions and $R(FoV)$ covering the remaining area.
We define $R(Buf)$ as an additional buffer region to cover VPs slightly deviated from the $R(FoV_f)$ and $R(FoV)$ which is likely to be outside the region covering at least 80\% of user VPs. Finally, $R(OoV)$ is the remaining area of $VM$ with the lowest viewing probability. Details of \textit{Pre-processing} steps are presented in Section~\ref{subsec:vp_aware_partitioning}.

In \emph{Partitioning} step, we run \textit{MNC} algorithm on the above four regions
separately. We create rectilinear polygon boundary around the blobs in these regions and partition them into non-overlapping tiles denoted as DT (Derived Tile) comprised of set of BTs (Basic Tiles).
Since we consider multiple blobs in VM frame separately, there can be overlaps between the derived tiles, if the considered blobs are closed to each other. Therefore we remove such overlaps during $R(FoV)$ and $R(Buf)$ partitioning. Details of \textit{Partitioning} steps are presented in Section~\ref{subsec:partitioning}.

In \textit{Post-processing} step, we further split DTs which are larger than FoV ($\ang{100}\times \ang{100}$) considering its distortion when projecting on to the Equirectangular (ERP) Frame. Finally, bit rate can be allocated to each DT considering the tile properties such as average pixel intensity, tile size and tile location.
Details of partitioning steps are presented in Section~\ref{subsec:post processing}.

Here on-wards, we denote a given user and video by $i$ and $j$ respectively, and no. of users in a video as $u_j$.
Pixel coordinates of video frames are denoted as (m,n), where $0\leq m < H$, $0\leq n < W$.


    
   

\begin{figure}[h]
  \centering
    \centering
    \includegraphics[width=\linewidth]{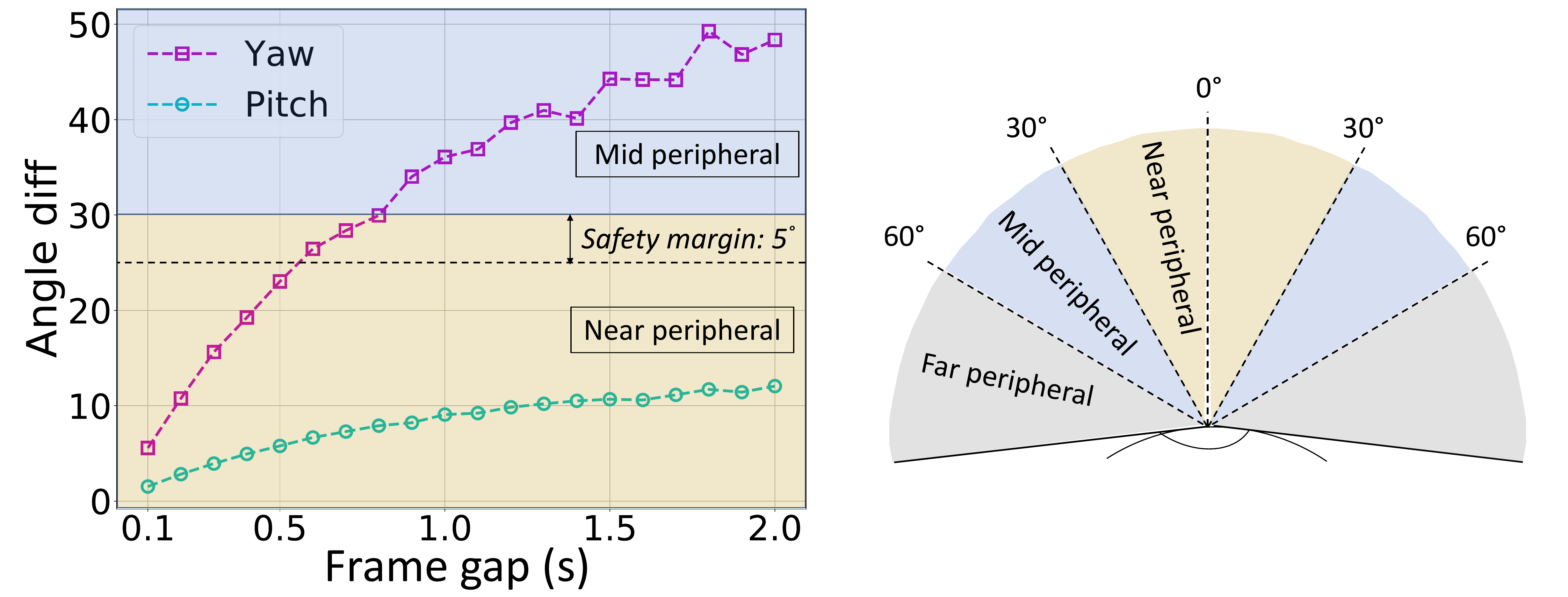}\vspace{-3mm}
    \caption{(left): Angle diff. vs the frame gap. (right): Peripheral vision of human eye}\vspace{-3mm}
    \label{fig:frame_sampling} 
\end{figure} 

\vspace{-2mm}
\subsection{Pre-processing}\label{subsec:vp_aware_partitioning}

We now describe the frame pre-processing pipeline  identify salient regions to be partitioned by the MNC algorithm as the output. 

\vspace{-1mm}
\subsubsection{Frame sampling (Fig.~\ref{fig:main_block}-a)}\label{subsubsec:frame sampling}

Frame rate for videos can vary between 24-60fps while 30fps is common in general.
It is not necessary to partition each and every frame in a video because; (i) it is safe to assume that FoV of the user is fixed at a certain position for certain period~\cite{nguyen2018your}, (ii) having different tile scheme for each frame can 
reduce compression gain in encoding, and (iii) based on the above two facts, running a partitioning algorithm on every frame adds unnecessary computational cost. 

To decide the most suitable frame gap, we analyse the relationship between the temporal and spatial user viewport behavior in light of the peripheral vision of human eye.
Fig.~\ref{fig:frame_sampling} shows the angular difference of yaw and pitch distribution against the sampling frame gaps from 0.1--2.0s by every 0.1s. 
Human vision can perceive high quality only at \textit{near peripheral region}, i.e. within \ang{30} range, ~\cite{bhise2011ergonomics,chaturvedi2019peripheral,strasburger2011peripheral}.
Based on that fact, we make a fair assumption that from a fixation point,
the user can view with almost the same visual quality at  maximum of \ang{30} without changing the fixation point.  
According to practical VP traces, up to 0.8s of frame gap can tolerate \ang{30} angle difference in Yaw direction. Including a safety margin of \ang{5}, we decide 0.5s as the frame gap to refresh the tile scheme without making a significant harm to the user perceived quality. Compared to 1s tile scheme refreshing period found in literature~\cite{xiao2017optile,ozcinar2019visual}, which can leads $(\geq\ang{35})$ angle difference, 0.5s gap can better adapt to FoV changes.
Also, it reduces the encoding overhead and time for processing if we are to consider every single frame for partitioning.


\vspace{-1mm}
\subsubsection{Viewport Map (VM) generation (Fig.~\ref{fig:main_block}-b).}
Despite many different approaches for generating visual attention maps, we leverage historical user VP traces to generate VMs. \cite{ozcinar2019visual} claims that 17 users are sufficient to create representative VP map. Therefore, we consider 20 users from each video to generate our VM frames.
First, given the centre of VP: $c_{i}$, of each user $i$ in $<yaw,pitch>$ angles, we create a binary map, $V_{i}$ according to Eq.~\ref{eq:vp map definitaion}. 
\vspace{-1mm}
\begin{equation}\label{eq:vp map definitaion}
    V_{i}(m,n)=
    \begin{cases}
      1, & \text{if}\ (m,n) \in F_{i} \\
      0, & \text{otherwise} 
    \end{cases}
\end{equation}

We assume a $\ang{100}\times\ang{100}$ FoV area ($F_i$) representing the FoV of the majority of commercially available HMDs~\cite{carlsson2020had}. 
We also 
project spherical coordinates of pixels $(x,y)$ to ERP format $(m,n)$ considering the geometrical distortion, creating more dispersed pixel distribution towards the upper and bottom parts of the frame (i.e., corresponding to the polar region of the spherical frame). 
We then generate $VM$ taking the average of all users, $u_j$. We generate normalized avg. Viewport Map, $VM^{norm}$, after histogram equalizing the $VM$ and dividing by maximum pixel value (=255) as in Eq.~\ref{eq:vm normalize}
\vspace{-1mm}
\begin{equation}\label{eq:vm normalize}
    VM^{norm}= \frac{1}{255}HIST(VM)
\end{equation}

\begin{figure}[h]
  \centering
    \begin{subfigure}{.49\columnwidth}
    \centering
    \includegraphics[width=\linewidth]{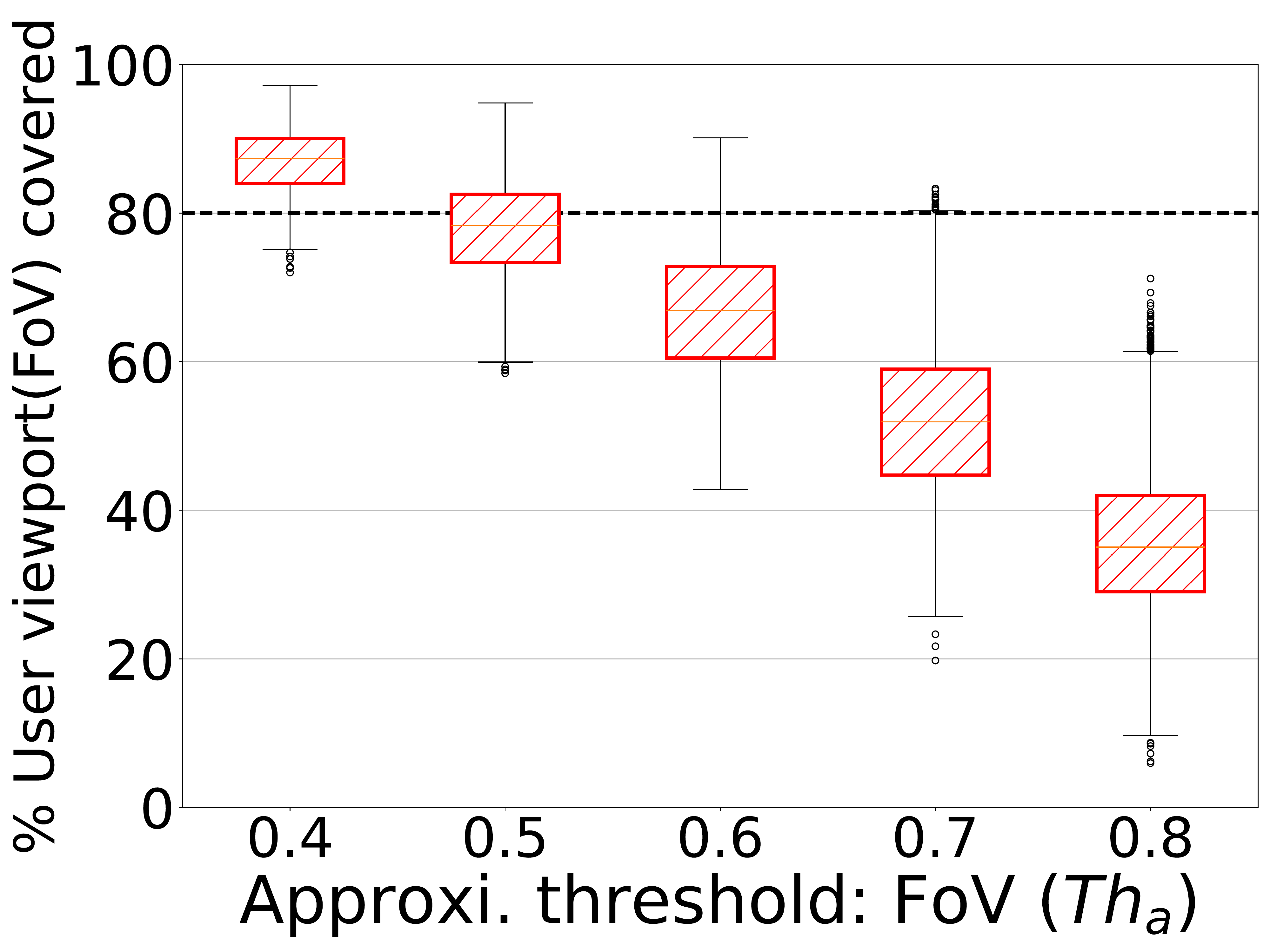}
    \vspace{-5mm}
    \caption{$Th_{a}$ validation}
    \label{fig:validate_app_threshold}
  \end{subfigure}
  \begin{subfigure}{.49\columnwidth}
    \centering
    \includegraphics[width=\linewidth]{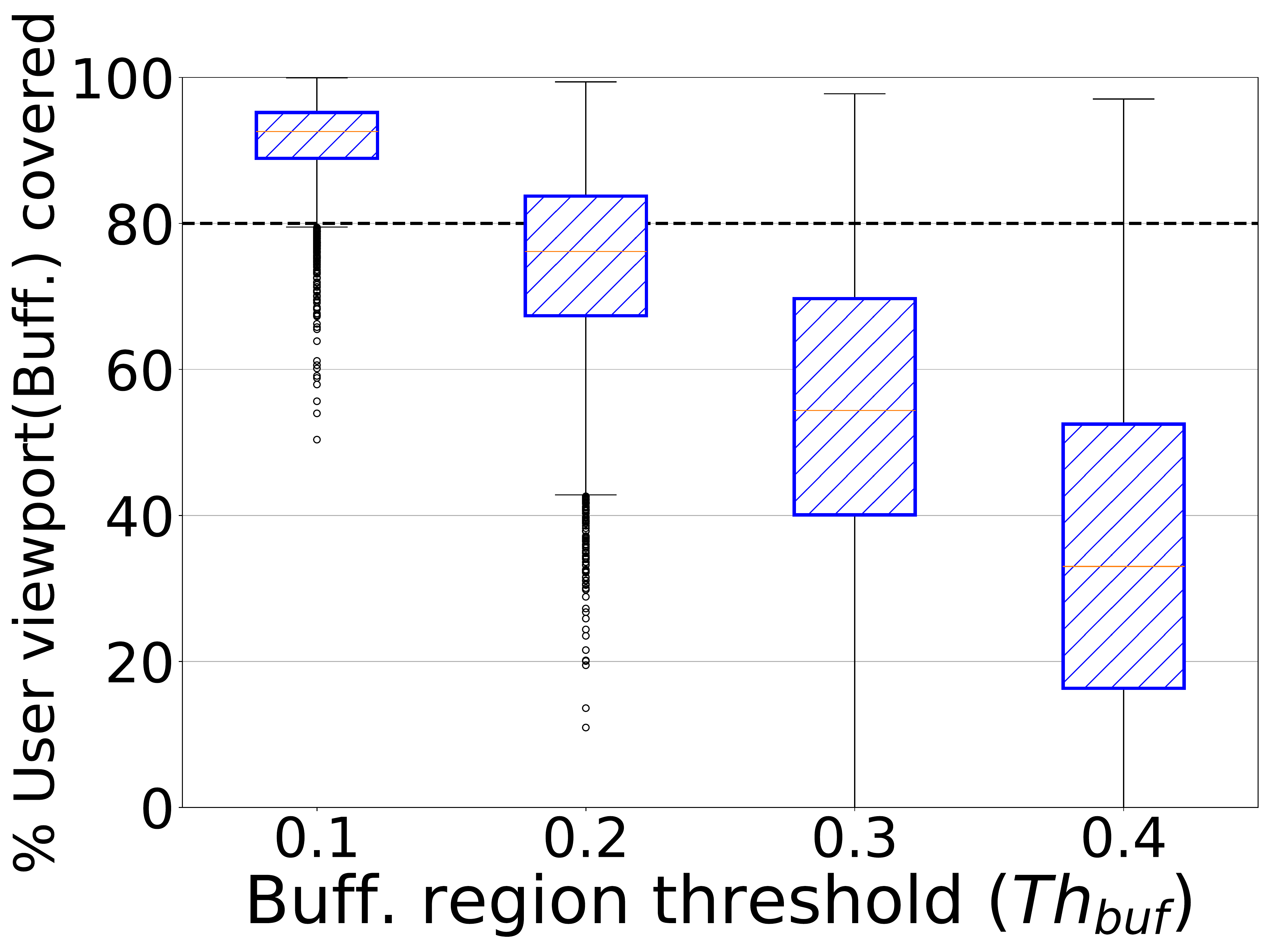}
    \vspace{-5mm}
    \caption{$Th_{buff.}$ validation}
    \label{fig:validate_buff_threshold}
  \end{subfigure}
\vspace{-3mm}
\caption{\% user viewport distribution on thresholded area under different threshold values for both approximate ($Th_a$) and buffer ($Th_{buff}$) region thresholding}
\label{fig:threshold param validation}
\end{figure}

\vspace{-2mm}
\subsubsection{Approximate Thresholding (Fig.~\ref{fig:main_block}-c)}\label{subsubsec:a_threshold}




   
    
    

    

\algblock{Input}{EndInput}
\algnotext{EndInput}
\algblock{Vaiable}{EndVaiable}
\algnotext{EndVaiable}
\newcommand{\Desc}[2]{\State \makebox[4em][l]{#1}#2}

\begin{algorithm}[h!]
\caption{Determine $Th_a$}\label{alg:th_a finding}

\begin{algorithmic}[1]
\Input
\Desc{$VM^{norm}$}{Normalized VP map}
\Desc{$\{V_i\}$}{Set of individual user VP maps $\forall i\in [1,u_j]$}
\EndInput

\Vaiable
\Desc{$\alpha$}{ a given $Th_a$ value s.t. $\alpha \in \{0.4,0.5,0.6,0.7\}$}
\Desc{$VM(\alpha)$}{binary map of $VM^{norm}$ after thresholding by $\alpha$}
\Desc{$I(\alpha)$}{Intersection map between $V_i$ and $VM(\alpha)$}
\Desc{$s_{i}(\alpha)$}{\% overlap between $V_i$ and $I(\alpha)$}
\Desc{$S(\alpha)$}{a set containing $s_{i}(\alpha)$ $\forall i\in [1,u_j]$}
\Desc{$s_{avg}(\alpha)$}{avg. of $s_{i}(\alpha)$ for all users $u_j$}
\EndVaiable

\For{$\alpha =$ 0.4 to 0.7, step = 0.1}
   
    \State   
    $VM(\alpha)=
    \begin{cases}
      1, & \text{if} \ VM^{norm}(m,n)\geq\alpha \\
      0, & \text{otherwise} 
    \end{cases}$
    
    \For{$i =$ 1 to $u_j$ }
        \State $I_{i}(\alpha)$ = $VM(\alpha) \cap V_{i}$ \Comment{get intersection map}
        
        \State $s_{i}(\alpha)=\frac{\sum\sum I_{i}(\alpha)}{\sum\sum V_{i}}\%$ \ $\forall m\in [0,H), \forall n\in [0,W), $ 
        
        
        \State $S(\alpha).add(s_{i}(\alpha))$ \Comment{store the \% overlap user $i$}
    \EndFor
    \State $s_{avg}(\alpha) = \frac{1}{u_j}\sum_{i=1}^{u_j} s_{i}(\alpha), \ \text{s.t.} \ s_{i}(\alpha) \in  S(\alpha)$

     \If{$s_{avg}(\alpha)<80\%$}\Comment{check for 80\% coverage}
        \If{$\alpha>0.4$}
            \State $Th_a^*=\alpha-0.1$
        \Else       \Comment{if none of the $Th_a$ satisfy 80\% coverage}
            \State $Th_a=0.4$
        \EndIf
        \State \textbf{return}  $Th_a$
     \EndIf
     
\EndFor
\State $Th_a=0.7$ \Comment{even if $\alpha=0.7$ satisfy the 80\% coverage}
\State \textbf{return} $Th_a$
\end{algorithmic}
\end{algorithm}


The objective of this step is to filter $R(FoV_f)$+$R(FoV)$ regions representing at least 80\% of the user VPs. To determine the appropriate threshold of $VM^{norm}$, we first calculated the overlap between individual user VPs and the thresholded region for a range of thresholds, $Th_a$ as shown in Fig.~\ref{fig:validate_app_threshold}. The results show that except 0.8, all other values can cover at least 80\% of FoV region at least for one frame. We also note that 80\% margin covers VPs of at least 17 users, out of 20 users, which is claimed to be the minimum number of user VPs needs to generate representative $VM$~\cite{ozcinar2019visual}.
By this analysis, we reduce our search space for appropriate threshold $Th_a$ between 0.4 to 0.7. Note that higher the $Th_a$, smoother the thresholded region boundary reducing the complexity of partitioning algorithm.


We present Algorithm~\ref{alg:th_a finding} to select highest possible threshold for each frame, $Th_a$.
First, we create binary map $VM(\alpha)$, by thresholding the $VM^{norm}$ using $Th_a=\alpha$ (line~12). Then we measure the \% of FoV overlap of each user VP ($V_i$) with $VM(\alpha)$ calculating an average value, $s_{avg}(\alpha)$ (line 13-17). Finally, if the $s_{avg}(\alpha)$ is no more giving 80\% coverage we stop further process and assign $Th_a$ with the previous $\alpha $ value. If none of the $Th_a$ satisfy 80\% FoV coverage, we select $Th_a=0.4$ (line~18-24). We apply $Th_a=\alpha^*$ on $VM^{norm}$ and denote the resulting frame with $R(FoV_f)$+$R(FoV)$ as $VM(\alpha^*)$.




        

\begin{algorithm}[h!]
\caption{Select blobs from $VM(\alpha^*)$}\label{alg:th_b finding}
\begin{algorithmic}[1]
\Input
\Desc{$VM(\alpha^*)$}{Approximate thresholded frame}
\EndInput

\Vaiable
\Desc{$b_{l}$}{ $l^{th}$ blob in $VM(\alpha^*)$, $l\in [1,l_{max}]$, $l_{max}$: maximum no. of blobs in $VM(\alpha^*)$}
\Desc{$z_{l}$}{size of $l^{th}$ blob}
\Desc{$B$}{set containing the all the blobs from $VM(\alpha^*)$}
\Desc{$B_{sel}$}{Set containing the selected blobs}
\Desc{$Z_{sel}$}{total size of a selected blobs from $B_{sel}$}
\Desc{$Z_{VM(\alpha^*)}$}{Total thresholded area of $VM(\alpha^*)$ }
\EndVaiable

\State $B\gets G(VM(\alpha^*))$ \Comment{get all the blobs to set $B$}
\State $B^{sort}\gets sort(B)$ : \text{(descending order of blob size)}
    
\State $Z_{sel} = 0$
\While{$l<l_{max}$}
    \State $B^{sel}.add(b_{l})$ \Comment{Add blobs to a set}
    \State $Z_{sel} = Z_{sel}+z_{l}$ \Comment{add blob size cumulatively}
    \If{$(Z_{sel}/Z_{VM(\alpha^*)})\%\geq 95$}\Comment{check for 95\% coverage}
        \State break
    \EndIf
    \State $l=l+1$ \Comment{increment the blob count}
\EndWhile
\State $VM^{{\alpha^*},b}\gets H(B^{sel})$ \Comment{create pixel map from selected blobs}
\State \textbf{return} $VM^{{\alpha^*},b}$
\end{algorithmic}
\end{algorithm}

\vspace{-2mm}
\subsubsection{Blob detection (Fig.~\ref{fig:main_block}-d)}\label{subsubsec:blob detection}
Due to the non-uniform dispersion of the user VP, $VM$ can contain multiple
blobs.
Aim of this step is to identify these blobs and exclude non-significant small blobs reducing the complexity of partitioning process. 
Without loss of generality, we select the blobs in $VM(\alpha^*)$, covering at least 95\% of  $R(FoV)$+$R(FoV_f)$. 
Algorithm~\ref{alg:th_b finding} summarizes the blob selection process given the $VM(\alpha^*)$ as the input. Firstly, the function $G(VM(\alpha^*))$ outputs all the blobs in  $VM(\alpha^*)$ frame as a set, $B$, followed by sorting in descending order according to the blob size (line 10-11). After that, we cumulatively sum up the blob size, starting from the largest one and stop the process when the total selected blobs size ($Z_{sel}$) exceeds 95\% of total thresholded area ($Z_{VM(\alpha^*)}$) in $VM(\alpha^*)$ (line 12-18). Finally, a map, $VM({\alpha^*},b)$, is created combining all the selected blobs using the function $H(B^{sel})$ 
(line 19).  



\vspace{-1mm}
\subsubsection{Finer thresholding (Fig.~\ref{fig:main_block}-e)}
To provide a higher quality for the most attractive region, in this step, 
we filter $R(FoV_f)$
from $VM({\alpha^*},b)$, defining \textit{Finer threshold}, $Th_f$. Without loss of generality we set $Th_f=0.9$ to identify the region $R(FoV_f)$  boundaries.
We expand this boundary to generate perfect rectangular polygon as we discussed in Section~\ref{subsec:partitioning}. We denote the finer thresholded  frame as $VM({\alpha^*},b,f)$, which is the input for \textit{MNC} algorithm for $R(FoV) + R(FoV_f)$ partitioning .

\vspace{-1mm}
\subsubsection{Buffer region thresholding and blob detection (Fig.~\ref{fig:main_block}-f and~\ref{fig:main_block}-g)} 
The objective of this step is to filter $R(buf)$ which covers slight variations of user VPs.
We extract $R(Buf)$ from
the area not covered by the Derived Tiles (DT) from $VM({\alpha^*},b,f)$ on the initial $VM$ which is denoted as $ VM^{buf}$ (Eq.~\ref{eq:buffer VM}). We use the same DT information to obtain corresponding buffer regions in $V_{i}$ (individual user viewports) namely, $V_{i}^{buf}$, as in Eq.~\ref{eq:buffer ind vp}.

\vspace{-2mm}
\begin{equation}\label{eq:buffer VM}
    VM^{buf}\gets VM\cap (VM^T({\alpha^*},b,f))^{'}
\end{equation}
\vspace{-4mm}
\begin{equation}\label{eq:buffer ind vp}
    V_{i}^{buf}\gets V_{i}\cap (VM^T({\alpha^*},b,f))^{'} \ \forall i\in [1,u_j]
\end{equation}

\noindent
where $VM^T({\alpha^*},b,f)$ and $(VM^T({\alpha^*},b,f))^{'}$ denotes the DT overlay on the $VM({\alpha^*},b,f)$ by \textit{MNC} algorithm and the remaining region on the frame respectively. 

In order to extract a suitable buffer region from $VM^{buf}$, we apply \textit{Buffer threshold} ($Th_{buf}$) and \textit{Blob detection} as the same way we followed in $Th_a$ finding and \textit{Blob detection} in approximate thresholding (cf. Section~\ref{subsubsec:a_threshold}~and~\ref{subsubsec:blob detection}). Hence, we compute \% user viewport ($V_{i}^{buf}$)  covered by the thresholded region from $VM^{buf}$, for the $Th_{buf}\in \{0.1,0.2,0.3,0.4\}$. 
Fig.~\ref{fig:validate_buff_threshold} shows that all threshold values can provide ($\geq 80\%$)\footnote{Covering corresponding $V_{i}^{buf}$ from at least 17 users out of 20} buffer viewport coverage, therefore, we dynamically select $Th_{buf}$ value from the above set. We apply Algorithm.~\ref{alg:th_a finding} simply changing the threshold values (line-5) and replacing $VM^{norm}$ (line-2~\&~12) and $V_{i}$ (line-3~\&~14) with $VM^{buf}$ and $V_{i}^{buf}$ respectively. We define the best $Th_{buf}=\beta^{*}$ and thresholded buffer frame as $VM_{i}^{buf}(\beta^{*})$. After that to exclude the non-significant smaller blob region, we apply Algorithm~\ref{alg:th_b finding} on $VM^{buf}(\beta^{*})$, by simply replacing $VM(\alpha^*)$ with $VM^{buf}(\beta^*)$ (line~2~\&~10). We denote blob filtered buffer frame as $VM^{buf}(\beta^*,b)$. 

\vspace{-1mm}
\subsubsection{OoV extraction}
The goal of this step is to extract $R(OoV)$ to derive low quality DT to satisfy any anomaly user VP. We filter out $R(OoV)$ removing the area covered by DT overlay on $VM({\alpha^*},b,f)+VM_{i}^{buf}(\beta^{*},b)$ area (similar to Eq.~\ref{eq:buffer VM}). No further pre-processing is applied to OoV region as no significant pixel value distribution is observed. We denote the OoV region as $VM^{oov}$.


\vspace{-2mm}
\subsection{Partitioning}\label{subsec:partitioning}
\model{} frame partitioning step runs the \textit{MNC} algorithm on $R(FoV_f)$+$R(FoV)$ , $R(Buf)$ and $R(OoV)$ regions separately to generate DTs. We start with creating a rectilinear polygon covering each blob followed by running basic \textit{MNC} algorithm in Section~\ref{subsec:basic_mnc}.

\begin{figure}[h]
    \centering
    \includegraphics[width=\columnwidth]{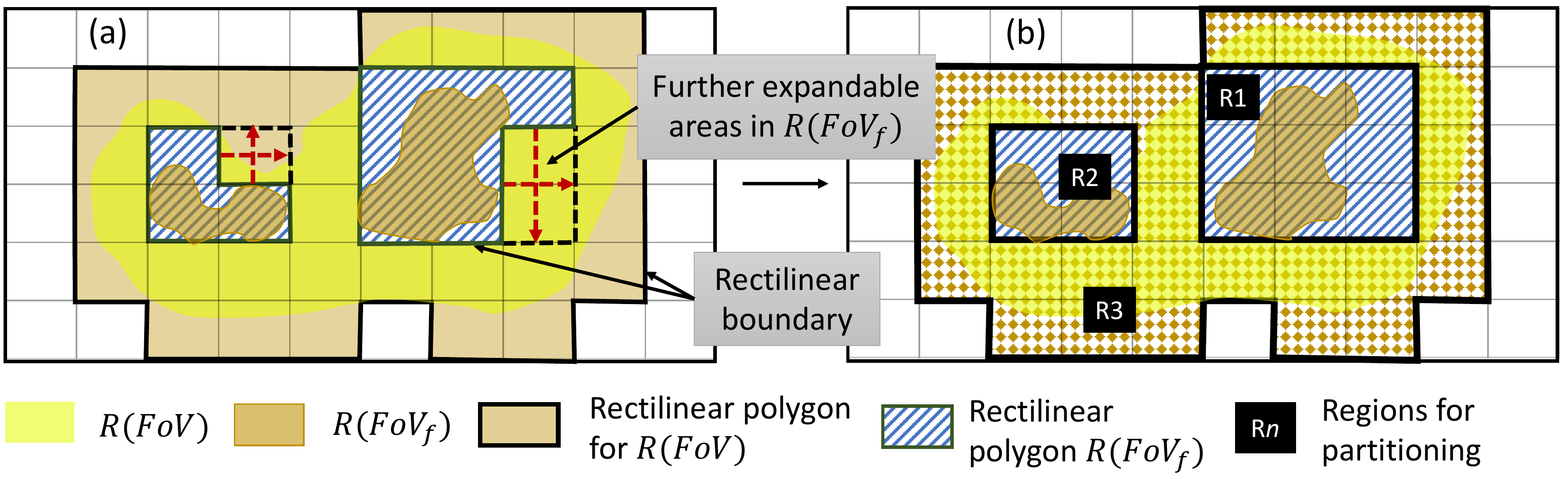}\vspace{-4mm}
    \caption{$R(FoV)+R(FoV_f)$ partitioning: (a)-before~\&~(b)-after expanding rectilinear polygon of $R(FoV_f)$}
    \label{fig:finer_boundary_transf}\vspace{-2mm}
\end{figure}

Firstly, for $R(FoV)$ and $R(FoV_f)$ partitioning (Fig.~\ref{fig:main_block}-h), we leverage $VM({\alpha^*},b,f)$. Fig.~\ref{fig:finer_boundary_transf} shows ($R(FoV)$ + $R(FoV_f)$) partitioning process. We expand the detected polygon in $R(FoV_f)$ (i.e., polygons in blue color) converting to a perfect rectangle.  
The boundary is extended to the minimum and maximum (m,n) locations as long as it does not exceed the $R(FoV)$ polygon boundary as shown in red color arrows. By this step, we make the partitioning process simpler and create an extra buffer for $R(FoV_f)$s to be encoded at higher quality. 
Fig.~\ref{fig:finer_boundary_transf}~(b) shows that polygons for $R(FoV)$ (e.g.,\textbf{R3}) is extracted removing all the polygons generated for $R(FoV_f)$ (e.g., \textbf{R1}, \textbf{R2}). 
Note that, extracting $R(FoV_f)$ as rectangles create holes in $R(FoV)$ region. 
Since, the basic \textit{MNC} algorithm is proposed for hole-free rectilinear polygons, we have added additional steps on top of \model{} \textit{MNC} implementation. In brief, when finding maximum independent chords, we take vertices of holes into account. Then, from the remaining vertices which have not connected with any independent chord, we draw extra chords to complete the partitioning. We avoid hovering any chord on the holes.

Secondly, taking $VM^{buf}(\beta^*,b)$ and $VM^{oov}$ frames \textit{MNC} algorithm partitions $R(Buf)$ and $R(OoV)$ respectively (Fig.~\ref{fig:main_block}-i and ~\ref{fig:main_block}-j) without any further processing on the polygons around the blobs detected. Finally, we see that, due to the close proximity of selected blobs certain DTs may overlap on each other. Given such two tiles, we remove the overlapped region only from the smaller DT ensuring the non-overlap DT coverage on the entire frame.

\vspace{-2mm}

\vspace{-5mm}
\begin{figure}[h]
    \centering
    \includegraphics[width=\columnwidth]{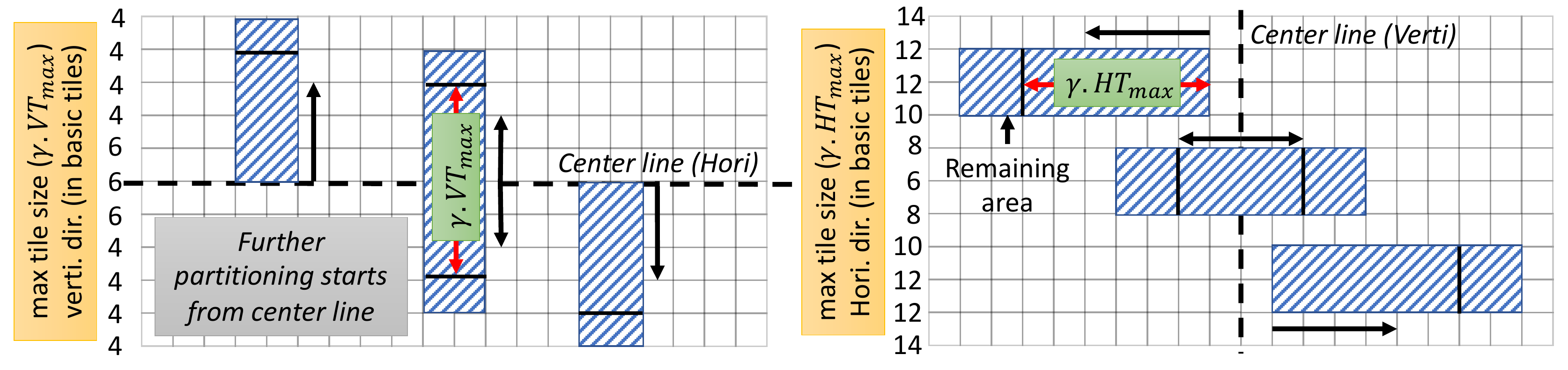}\vspace{-2mm}
    \caption{Further partitioning mechanism and maximum allowable tile size in horizontal and vertical direction based on the vertical position of center of a given tile.}\vspace{-3mm}
    \label{fig:further partitioning}
\end{figure}

\vspace{-4mm}
\subsection{Post-processing}\label{subsec:post processing}

\subsubsection{Further partitioning of bigger DTs}
We further partition DTs beyond certain limit of the size, in order to reduce the pixel redundancy. Since, we consider polygon boundary for the partitioning, we may encounter DTs even bigger than FoV size horizontally/vertically or in both. Therefore, any slight overlap with such tile incurs large pixel redundancy. In this process, we first define maximum allowable DT size considering the FoV distortion variation according to its vertical position. For example, viewport located towards polar region allows to have a larger DT size as the corresponding FoV on the equirectangular frame spread in a larger region compared to the equator. Hence, considering the \# of overlapped tiles with the distorted FoV maps on the equirectangular frame, we define 
maximum allowable DT size in both vertical ($VT_{max}$) and horizontal ($HT_{max}$) directions  which is shown in Fig.~\ref{fig:further partitioning} y-axis.

Note that further reducing $VT_{max}$ and $HT_{max}$ support decreasing redundant data transmission as the DT size becomes small, nonetheless incurs high encoding overhead. To see the impact, we take $\gamma .VT_{max}$ and $\gamma .HT_{max}$ where $\gamma \in[0,1]$. We set $\gamma=\{0.25,0.5,1.0\}$ in our experiments. Decreasing the $\gamma$ results in smaller tiles. After detecting larger DTs, we start partitioning outwards from the center lines as in example tile in Fig.~\ref{fig:further partitioning}. 
The reason is that majority of the user VP concentrated around the center of the frame~\cite{lo2017360,corbillon2017360}. Therefore, 
to reduce potential quality changes within the tile, we keep those DTs near center lines non-splitted as much as possible.

Finally, quality allocated tile scheme an be achieved as in Fig.~\ref{fig:main_block}-l considering the multiple properties of DTs such as pixel intensity, size and location of the tile. In \model{} we do not implement proper bit-rate allocation scheme and keep it as a future work. Further interactions with bit-rate allocation is discussed in Section~\ref{sec:discuss}.

\vspace{-2mm}
\section{Evaluation setup}

\subsubsection{Dataset}\label{subsubsec:dataset}
We develop and validate algorithms in \model{} leveraging VP traces collected from 30 videos from three different datasets~\cite{lo2017360,wu2017dataset,nasrabadi2019taxonomy}. All videos are in 60s duration with 30fps. VP center is denoted by $<yaw,pitch>$ angle. 
Each video has 30 users and we take VP traces from randomly selected 20 users to develop tile schemes using \model{} and the remaining 10 user VPs to validate the performance of \model{}.
The selected videos represents \ang{360} video categorization proposed in~\cite{carlsson2020had} and are in different genres such as sports, documentary, stage performance etc. generalizing the content of the videos. 

\vspace{-1mm}
\subsubsection{Hardware and software setup}
We implement \model{} architecture using Python, which consists of 4500 lines of code, on MacOS--intel Core i9 2.3GHz single core CPU. We use \textsf{Networkx-2.4} package for implementing the \textit{MNC} algorithm\footnote{For bipartite graph generation for searching maximum independent chords in a given polygon (cf. Section~\ref{subsec:basic_mnc})}. Videos in HD ($1920\times 1080$) and 4K ($3840\time 2160$) resolution are encoded using \textsf{FFmpeg}-4.1 in H265 (HEVC)  provided by \textsf{libx265} at default Quantization Parameter (QP)=28 with motion constrained tiling. 


\vspace{-1mm}
\subsubsection{Evaluation metrics and comparison benchmarks}
We compare \model{} with 3 fixed tile configurations $4\times 6$~\cite{qian2018flare}, $6\times 6$~\cite{he2018rubiks}, and $10\times 20$~\cite{lo2017360}. To evaluate \model{} with viewport-aware streaming, we use two metrics. \textit{i)}~\% Pixel redundancy before compression: extra pixels in selected DTs (Derived Tiles) but not overlapped with the user FoV using Eq.~\ref{eq:pixel redundancy}. Higher the value of pixel redundancy, num. of pixel level operations will increase in video coding at the servers and rendering at the client devices. 
\textit{ii)}~Downlink (DL) data volume: Data transmitted  by selected tiles in individual user VPs, which impacts the bandwidth saving. 
Total num. of DTs covering the entire frame when $\gamma = 1$ and $0.5$ is near similar to the fixed tile $4\times 6$ and $6\times 6$ cfgs. respectively. We compare \model{}'s relative gain to fixed tiling using the above metrics, and denote two cfgs. as \textbf{C1}: $\gamma=1$ to $4\times 6$ and \textbf{C2}: $\gamma=0.5$ to $6\times 6$.

\vspace{-1mm}
\section{Results}

We evaluate \model{} considering the DT (derived tile) distribution on the video frame and individual user VP, \% of pixel redundancy before compression, DL data volume  and frame processing time.

\vspace{-2mm}
\subsection{Distribution of DTs on the frame and overlap with user VP}\label{subsec:DT dist}

We first analyse the no. of DTs generated by \model{} on each region: $R(FoV_f)$, $R(FoV)$, $R(Buf)$ and $R(OoV)$. 
Since, the $\gamma$ value control the maximum allowable DT size (cf. Section.~\ref{subsec:post processing}), we vary the $\gamma$ 
to see the corresponding variation of no. of DT tiles and their size in BTs on each region.
Table.~\ref{table:DT distribution} reports the averaged results for all the frames in 30 videos.
For each $\gamma$ value, around 31\% of DTs on the entire frame covers $R(FoV)$, however, the avg. tile size is 37.3\% lower than the DTs in $R(FoV_f)$. This is due to the region expansion of $R(FoV_f)$ during \textit{Partitioning} step (cf. Section~\ref{subsec:partitioning}) and the remaining regions on the $VM(\alpha^*,b,f)$ covered by $R(FoV)$ are smaller patches surrounding the $R(FoV_f)$.
Second largest tiles are derived in $R(OoV)$ area as the maximum allowable tile size is higher near the upper and bottom region of the ERP frame.

\vspace{-2mm}
\begin{table}[h]
\small{
    \caption{DT distribution in different regions: no. of tiles (\#~T )  and avg. tile size in BTs (S) variation by $\gamma$}\vspace{-3mm}
    \label{table:DT distribution}
    \begin{tabular}{cccccccccc}
        \toprule
        \footnotesize{\textbf{$\gamma$}}& 
        \multicolumn{2}{c}{\footnotesize{\textbf{$R(FoV_f)$}}} & 
        \multicolumn{2}{c}{\footnotesize{\textbf{$R(FoV)$}}} & 
        \multicolumn{2}{c}{\footnotesize{\textbf{$R(Buf)$}}} & 
        \multicolumn{2}{c}{\footnotesize{\textbf{$R(OoV)$}}} & 
        \footnotesize{\textbf{Total}} \\
        & \#  T & S  & \#  T & S & \#  T & S & \#  T & S & \textbf{\#  T}\\        
         
        \midrule
        \footnotesize{\textbf{0.25}} & 18 & 3.2  & 19 & 2.8 & 11 & 3.0 & 16 & 3.6 & 64\\
        \footnotesize{\textbf{0.50}} & 8 & 6.8 & 13 & 4.0 & 8 & 4.0& 11 & 5.4 & 40 \\
        \footnotesize{\textbf{1.00}} & 4 & 13.4 & 9 & 5.6 & 6 &5.6 & 9 & 6.4 & 28 \\ 
        \bottomrule
    \end{tabular}
    }
\end{table}



\begin{figure}[t]
    \centering
    \vspace{-2mm}
    \includegraphics[width=\columnwidth]{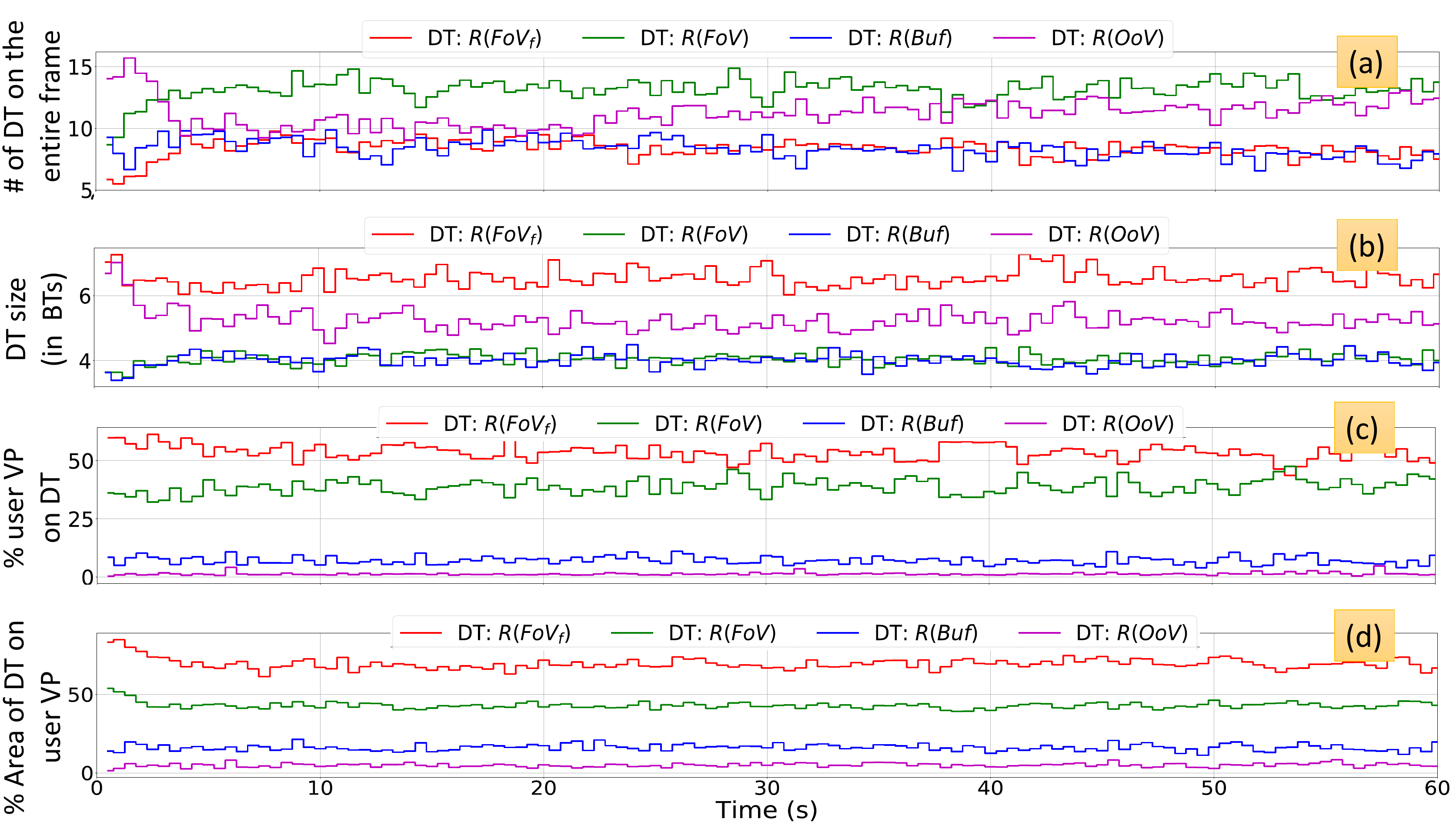}\vspace{-2mm}
    \caption{Temporal variation of DT at $\gamma=0.5$: (a) no. of DT on the entire frame, (b) size of DT, (c) \% user VP on DT (d)\% area of each DT overlap with user VP}
    \label{fig:temporal variation DT on frame and FOV}\vspace{-3mm}
\end{figure}

Fig.~\ref{fig:temporal variation DT on frame and FOV} shows the temporal variation of DT distribution for the entire video duration for $\gamma = 0.5$. Fig.~\ref{fig:temporal variation DT on frame and FOV}(a)~and~Fig.~\ref{fig:temporal variation DT on frame and FOV}(b) show that DT distribution becomes stable within first 5s. For example, in Fig.~\ref{fig:temporal variation DT on frame and FOV}(a), no. of DTs in $R(OoV)$ start decreasing from 15 to 10, in contrast DTs in $R(FoV_f)$ and $R(FoV)$ starts increasing from 5 to 10 and 10 to 14 respectively. 
Fig.~\ref{fig:temporal variation DT on frame and FOV}(b) illustrates that, DT size of $R(OoV)$ decreases from 7 to 5 whereas DTs in $R(FoV_f)$ and $R(FoV)$ keep nearly the constant tile size at 6 and 4. 
These observations conclude that within first 5s, \model{} generates large $R(FoV_f)$ and high no. of $R(OoV)$ DTs as the user VPs are concentric to a certain area. As the user VP start spreading on the frame, \model{} generates more $R(FoV)$ and $R(FoV_f)$ DTs because, no. of blobs with high visual attention have increased, reducing the $R(OoV)$.

Fig.~\ref{fig:temporal variation DT on frame and FOV}(c) shows \% user VP overlap with DTs from different regions. We see that, more than 50\% and 30\% of individual user VP overlaps with DTs from $R(FoV_f)$ and $R(FoV)$ enabling content providers to allocate high quality for DTs in the user VPs. Fig.~\ref{fig:temporal variation DT on frame and FOV}(d) shows the proportion of each DT area overlapped with the user VP. 
Also, starting from 80\%, avg. overlapped proportion of DT tiles on $R(FoV_f)$ decrease  to 70\%, 
as at the beginning, $R(FoV_f)$ DTs can provide finer boundary to the individual user VP, but slightly fails with VP dispersion. Only 48\% of the area of $R(FoV)$ DTs overlap with user VPs. The reason is many $R(FoV)$ DTs cover boundary of the high visual attention areas which results in low overlap with VPs.

\begin{figure}[t]
    \centering \vspace{-2mm}
    \includegraphics[width=\columnwidth]{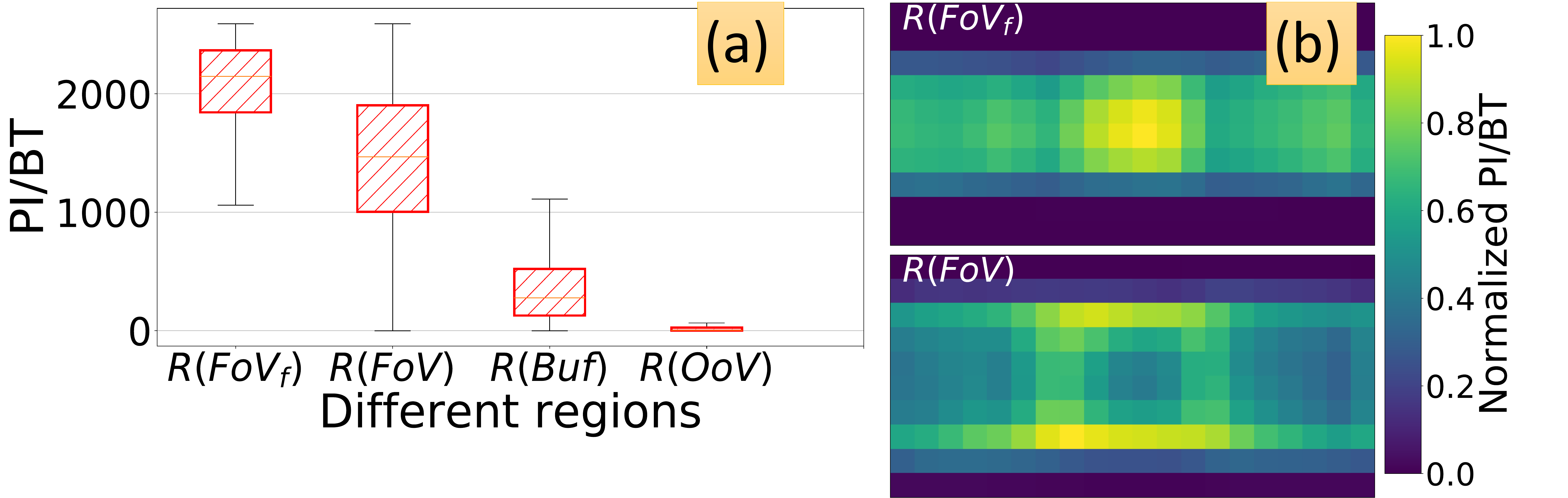}\vspace{-3mm}
    \caption{Pixel intensity distribution: (a)~Pixel Intensity per Basic Tiles (PI/BT) for the 4 regions, (b)~Spatial distribution of PI/BT  of $R(FoV_f)$ and $R(FoV)$}
    \label{fig:pixel intensity regions}\vspace{-3mm}
\end{figure}

Fig.~\ref{fig:pixel intensity regions}(a)shows the \textit{Total Pixel Intensity per Basic Tile} (PI/BT) of the DTs on 4 different regions on the frame, which corresponds to the visual attention level.
Overall, $R(FoV_f)$ and $R(FoV)$ attain more than 1000 PI/BT whereas the majority of DTs in both $R(Buf)$ and $R(OoV)$ have ($\leq 500$) PI/BT showing the effectiveness of \model{} threshold selection in \textit{Pre-processing} step. Moreover, Fig.~\ref{fig:finer fov spatial}(b) illustrates the spatial distribution of PI/BT values of $R(FoV_f)$ and $R(FoV)$. Values are normalized separately for the two regions for the clarity of presentation. The pre-defined $Th_f$ (=0.9) in \model{} is able to derive the majority of $R(FoV_f)$ tiles at the center of the frame, at where the user viewports are concentrated in general~\cite{wu2017dataset,corbillon2017360}. In the meanwhile, DTs in $R(FoV)$ covers the surrounding regions of $R(FoV_f)$ acting as a high quality buffer to $R(FoV_f)$.

\emph{Overall, DT distribution on identified 4 regions has unique properties in different aspects such as no. of tiles, size, overlap with user VP and pixel intensity levels. Proper understanding of these properties enables content providers to treat DTs at the servers, adaptively changing their quality levels in order to provide high QoE to the users.}



\begin{figure}[t]
  \centering 
    \begin{subfigure}{.49\columnwidth}
    \centering
    \includegraphics[width=\linewidth]{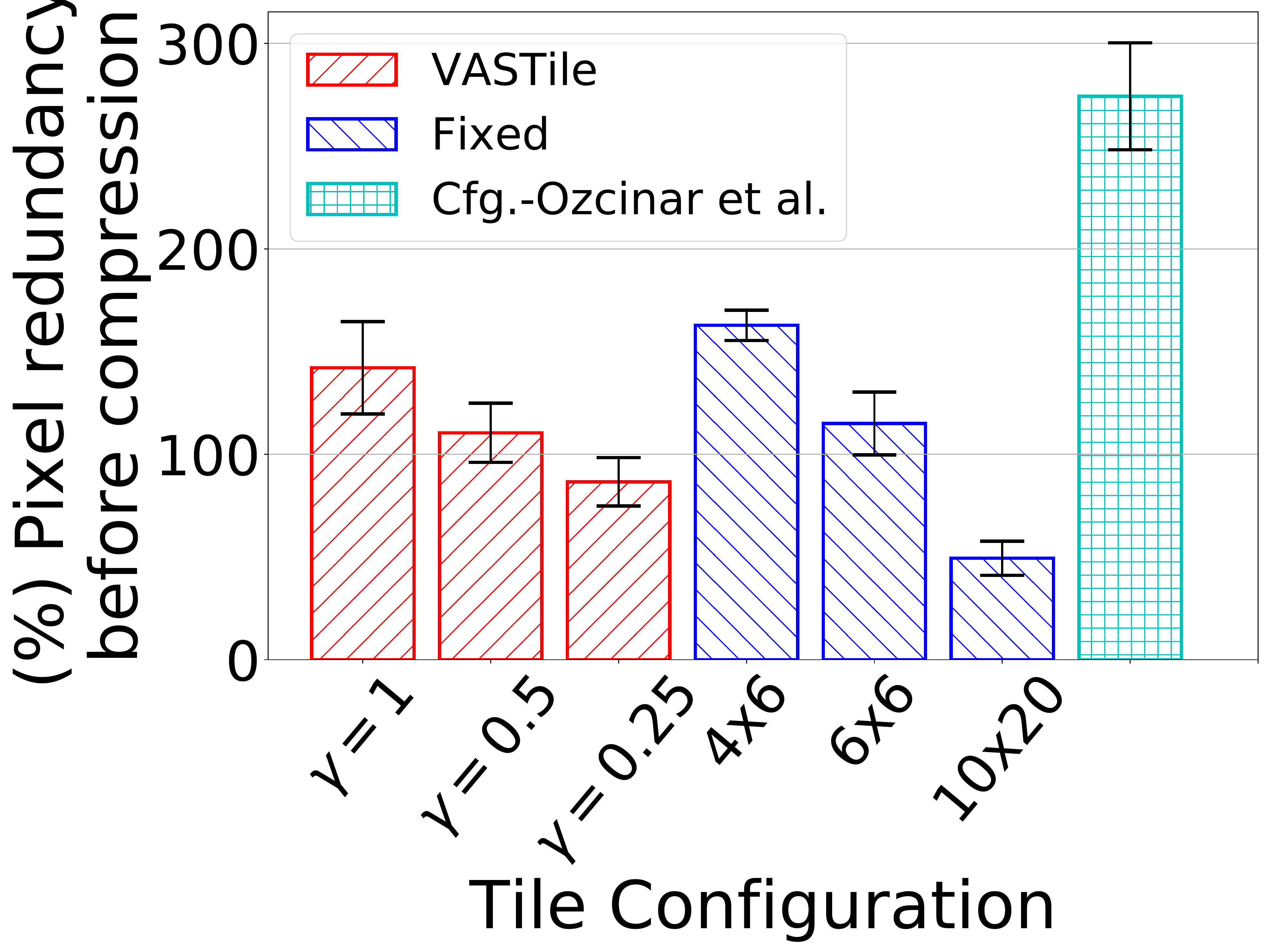}
    \caption{Absolute pixel redundancy for the entire video}
    \label{fig:abs pixel red}
  \end{subfigure}
  \begin{subfigure}{.49\columnwidth}
    \centering
    \includegraphics[width=\linewidth]{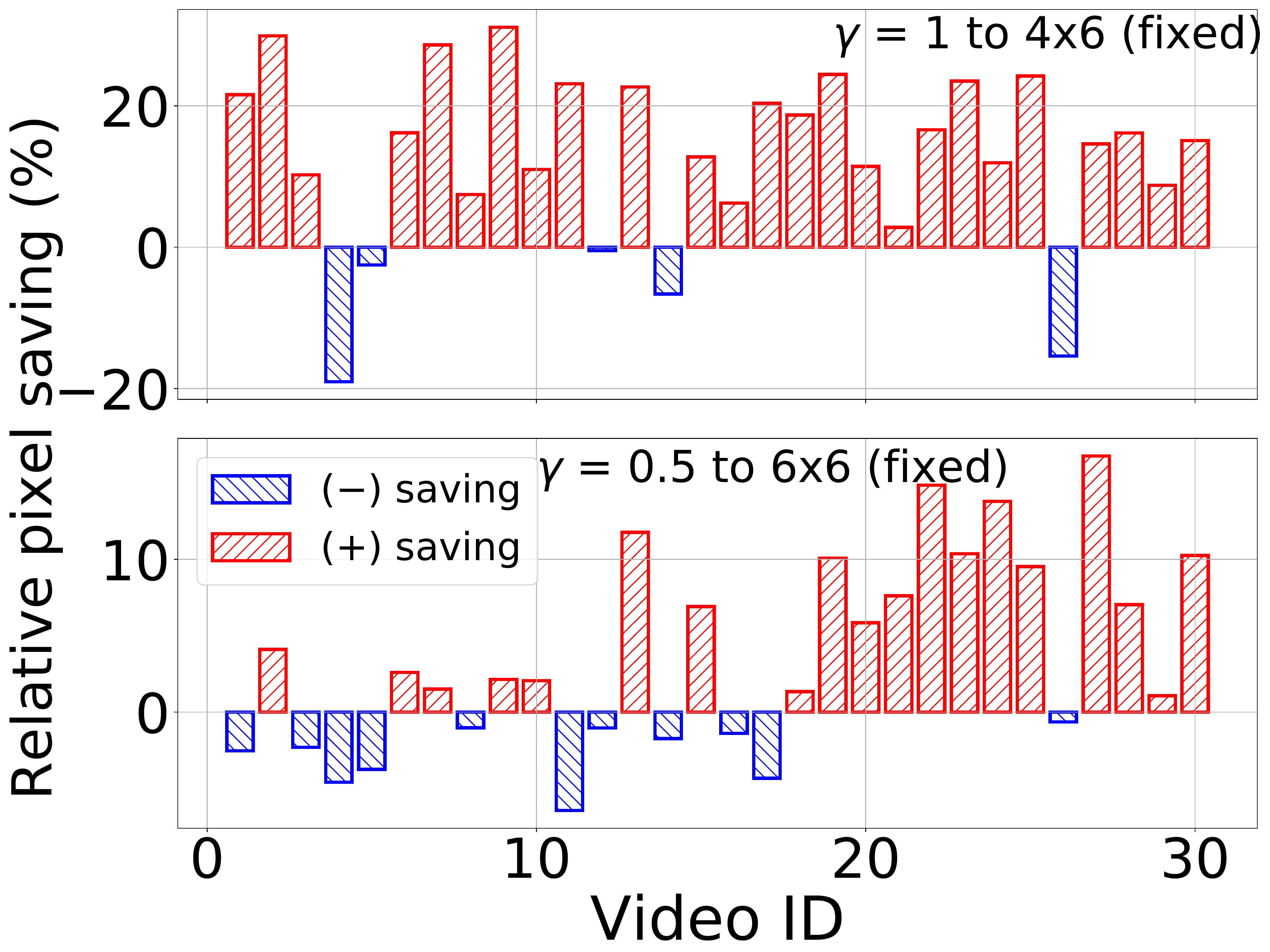}
    \caption{Relative pixel redundancy saving :\model{} to fixed cfgs.}\vspace{-2mm}
    \label{fig:rel pixel redund}
  \end{subfigure}

\vspace{-3mm}
\caption{Pixel redundancy before compression and relative gain achived by \model{} compared to fixed cfgs.}\vspace{-4mm}
\label{fig:threshold param validation}
\end{figure}

\vspace{-2mm}
\subsection{\% Pixel redundancy before compression}\label{subsec:pixel redund}

We measure \textit{\% pixel redundancy before compression} for the entire video using the Eq.~\ref{eq:pixel redundancy}. Fig.~\ref{fig:abs pixel red} shows the avg. results for all the videos. We see that when $\gamma=1$ ($\gamma=0.5$), \model{} redundant cover is 142 (110)\% which is a 20(5)\% reduction compared to fixed configuration $4\times 6$ ($6\times 6$). 
Such high redundancy is due to the partial overlap of the user VP by the DTs towards the boundary of the FoV. 
We further compare the best basic tile configuration provided by Ozcinar \textit{et al.}~\cite{ozcinar2019visual}.
Compared to \model{}:~$\gamma=1$, this approach costs additional 137\% of pixel redundancy mainly due to the large polar region tiles in their tile scheme.

Fig.~\ref{fig:rel pixel redund} shows relative saving of redundant pixels before compression in \textbf{C1} and \textbf{C2} cases for each individual video.
We measure an avg. of 12.8\% pixel saving and a maximum of 31.1\% in C1 ($\gamma=1$ to $4\times 6$). However the same results for C2 ($\gamma=0.5$ to $5\times 8$) are 3.7\% (avg.) and 16.8\% (max). The lower performance in C2 is due to the, reduced size of the  fixed cfg. tiles. 
This is the same reason for negative saving (i.e., more pixel redundancy in \model{}) in certain videos, nevertheless the negative saving is less significant compared to total positive savings. We can not expect the similar patterns in both graphs because, position of the tiles vary when changing the tile cfg. relative to the same user VP, affecting the pixel redundancy.

\vspace{-2mm}
\subsection{DL data volume}\label{subsec:dl_vol}

\begin{figure}[h]
  \centering
   \begin{subfigure}{0.49\columnwidth}
    \centering
    \includegraphics[width=\linewidth]{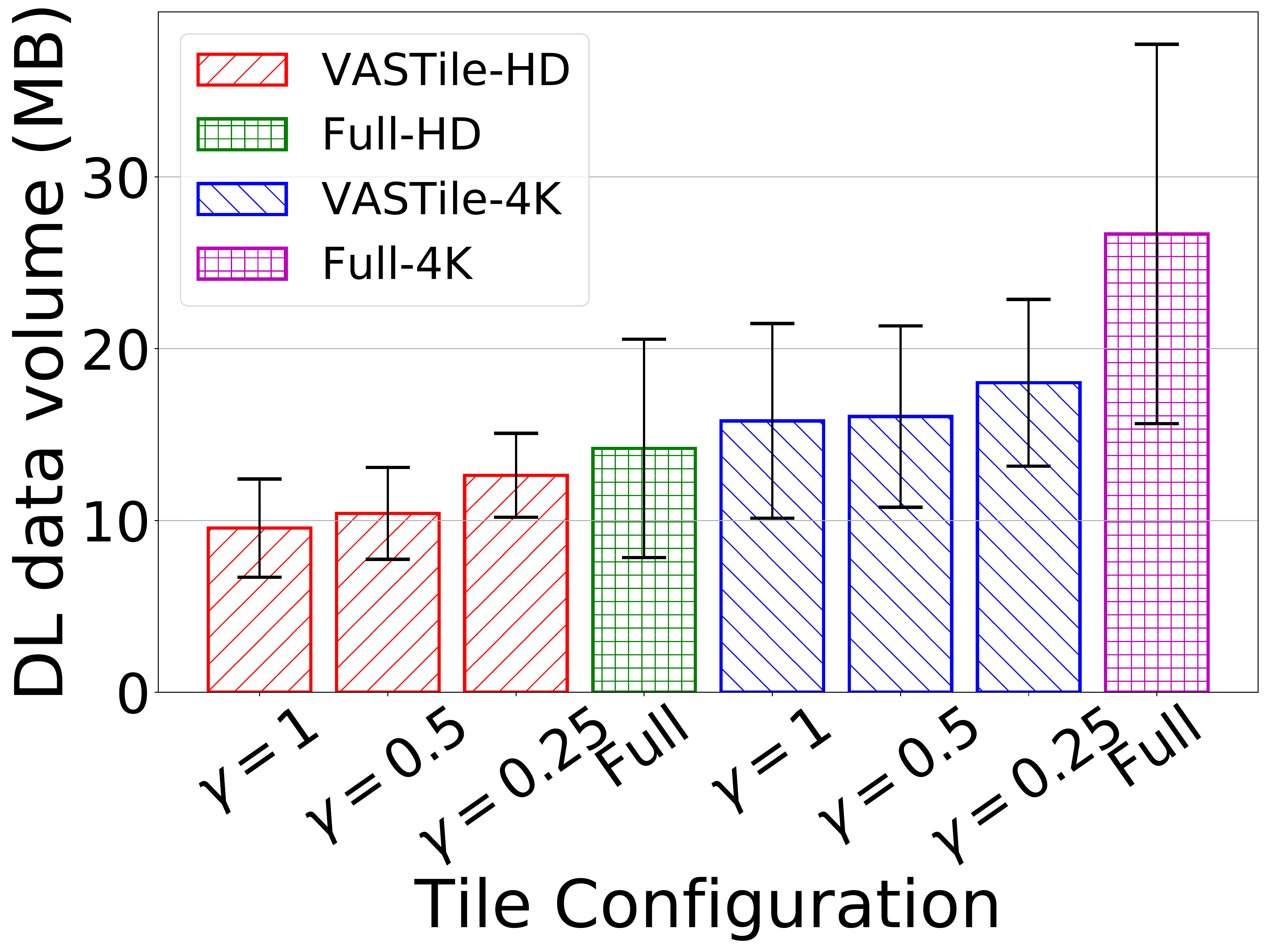}\vspace{-1mm}
    \caption{\model{} with viewport-aware vs Full-frame streaming}
    \label{fig:vastile vs full frame}
  \end{subfigure}
  \begin{subfigure}{0.49\columnwidth}
    \centering
    \includegraphics[width=\linewidth]{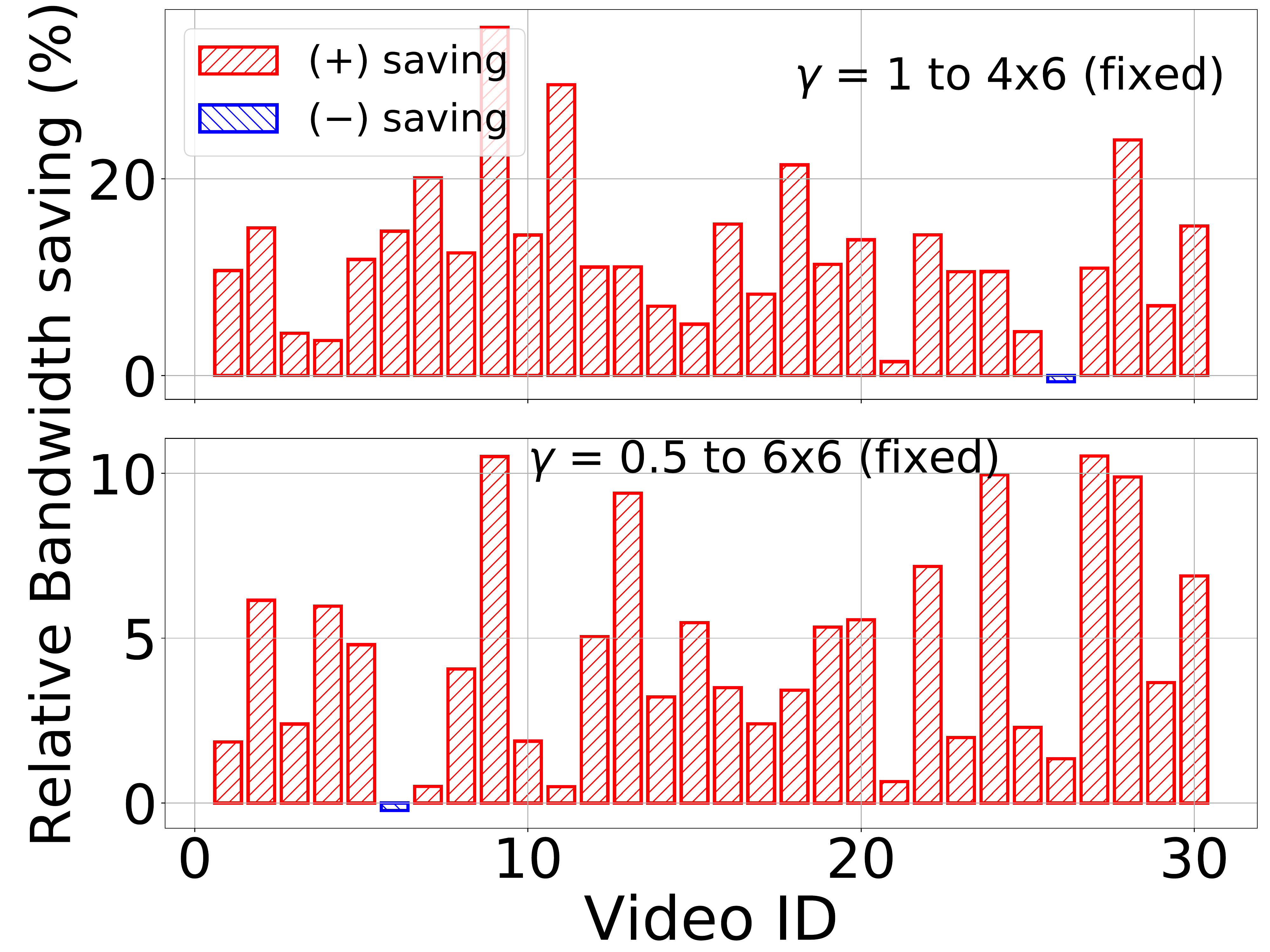}
    \caption{\model{} vs fixed tile cfgs for HD resolution}\vspace{-2mm}
    \label{fig:vastile vs fix}
  \end{subfigure}

\vspace{-3mm}
\caption{Comparison of \model{} with Full frame and Fixed tile cfgs with viewport-aware streaming}
\label{fig:threshold param validation}
\end{figure}

We compare the DL data volume in viewport-aware streaming using \model{} with streaming the entire video frame for both HD and 4K videos.  Fig.~\ref{fig:vastile vs full frame} shows that, compared to Full frame streaming scenario, which is utilized by commercial content providers such as YouTube and Facebook, \model{} can save in avg. 32.6 (40.8)\% of bandwidth in HD (4K) videos. When decreasing the $\gamma$, bandwidth saving reduces to 10.8(32.3)\%. The reason is, decrease of $\gamma$ affects increase in the no. of tiles, and thereby more encoding overhead. 

We further measure the bandwidth saving by \model{} with compared to the existing fix tile cfgs. for each video in Fig.~\ref{fig:vastile vs fix} similar to the Fig.~\ref{fig:rel pixel redund}. As in the Section.~\ref{subsec:pixel redund}, \textbf{C1} shows more gain, which is 12\% in avg. and 35.4\% in maximum. The similar measures for \textbf{C2} is 4.5\% (avg.) and 10.5\% (max). Almost all videos show positive bandwidth saving as the compression can boost up the \model{} performance despite negative relative pixel redundancy (cf. Fig.~\ref{fig:rel pixel redund}). 

\emph{Analysis of pixel redundancy before compression and DL data volume after encoding show that \model{} outperforms fixed tile cfgs. in general, while achieving significant gains with certain videos. Reduced pixel redundancy benefits in decreasing pixel level operations such as in encoding at the content servers and frame rendering at the client side. Moreover, bandwidth savings compared to full frame streaming enables network providers to relive the high strain on the network by current \ang{360} video streaming scenario.}

\vspace{-2mm}



\vspace{-8mm}
\subsection{Processing time of \model{}}\label{subsec:process time}
\begin{figure}[t]
    \centering\vspace{-2mm}
    \includegraphics[width=\columnwidth]{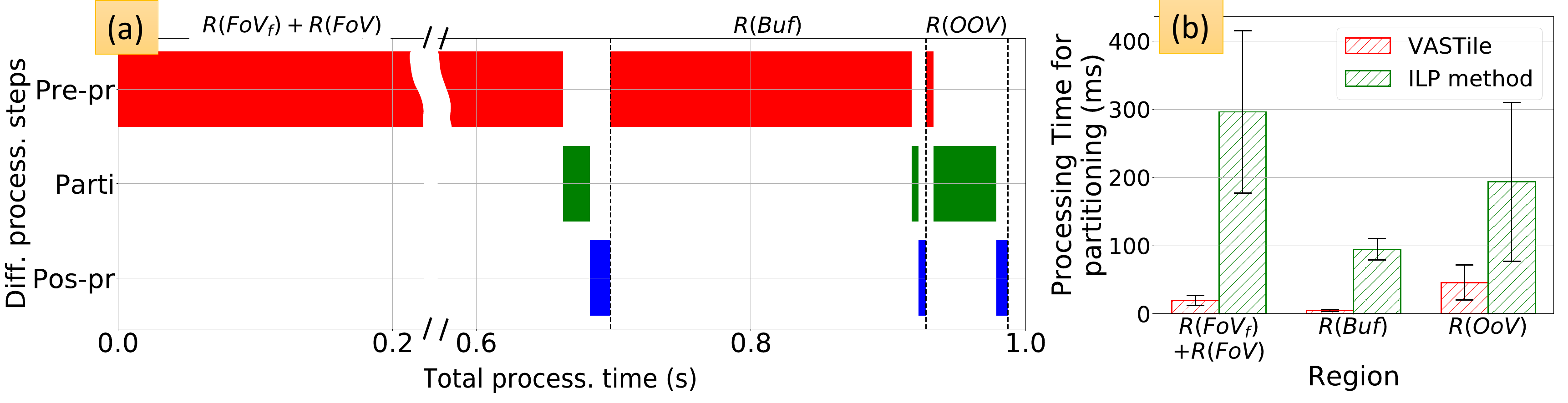}\vspace{-3mm}
    \caption{Time complexity: (a)~Dissection of VASTile process in time, (b)~Comparison with ILP  for \textit{Partitioning} step}
    \label{fig:time analysis}\vspace{-2mm}
\end{figure}

We measure the end-to-end  processing time for \model{} including \textit{Pre-processing}, \textit{Partitioning} and \textit{Partitioning} steps for the four main regions
as shown in Fig.~\ref{fig:time analysis}(a). In a gist, \model{} can provide a suitable tile scheme within 0.98($\pm 0.11$)s of avg. processing time revealing its scalablity for large scale video datasets. Pre-processing step in $R(FoV_f)$+$R(FoV)$ and $R(Buf)$ demands higher processing time due to the semi-automated selection of $Th_a$ and $Th_{buf}$. Comparing the ILP based approach~\cite{xiao2017optile} which takes 7--10s to process one frame on single core CPU(3.3GHz),
\model{} can reduce the processing time by 85--90\%. 


Fig.~\ref{fig:time analysis}(b) shows a comparison of processing time of \textit{Partitioning} step between \model{} and modified ILP based method from~\cite{xiao2017optile}. Since we consider the minimum number of tiles to cover the video frame, we modify the cost function  in ILP method to reduce the no. of tiles whilst keeping \textit{Pre/Post-processing} steps as the same.
Overall, \textit{Partitioning} time in \model{} is 513.2 ms \textbf{less} than ILP. 


\emph{Time for processing a suitable tile scheme by \model{} is less than 1s showing its potential scalability even for real time streaming. Leveraging an ILP/exhaustive type of approach need additional 0.5s reducing the scalability of the whole process.  }

\vspace{-2mm}
\section{discussion and limitations}\label{sec:discuss}

\subsubsection{Server storage}
\model{} does not encode all possible tiles when selecting a suitable tile scheme excluding compression related parameters such as motion vector distribution~\cite{xiao2017optile}. Therefore, the solution is sub-optimal reducing server storage optimization opportunities. Incorporating  compression awareness to \model{} should not obstruct the current low processing time. Therefore, we plan to input compression aware parameters to \model{}, utilizing Machine Learning based approach circumventing time consuming steps such as all possible tile scheme compression.

\vspace{-1mm}
\subsubsection{Support for DASH protocol}
True benefits of \model{} can be achieved with a  systematic implementation of dynamic bitrate/QP allocation to change the tile quality, in contrast to the constant QP value used at this stage. Compared to fixed tile cfgs, DTs from \model{} are aware of the viewport distribution on the video frame. Therefore, when adapting to the existing DASH protocols, we can have more control on adjusting the suitable quality affecting parameters (i.e., change the bitrate/QP levels based on the pixel intensity of the tiles on the $VM$s). 
We keep DASH implementation of \model{} and in-detail user QoE  analysis as our future work.


\vspace{-2mm}
\section{conclusion and future works}

In this paper, we proposed \model{}, a viewport aware adaptive \ang{360} video frame partitioning mechanism which derived a suitable tile scheme with variable sized tiles after identifying visually attractive regions on the frame.
We leveraged a computational geometric approach to find the minimal non-overlapping cover on these regions and combined the derived tiles on separate regions to cover the entire frame. We evaluated \model{} in-terms of derived tile distribution both on the video frame and individual user FoVs, saving of pixel redundancy before compression, bandwidth savings and scalability based on the processing time. Our results showed \model{} outperforms current \ang{360} video streaming scenario with full frame streaming and recently proposed fixed size tiling schemes.

In future work, we aim to investigate on DASH implementation on \model{} and related QoE aspects. Moreover, we aim to utilize content based saliency maps supporting content providers to encode the frames at a prior stage even without having user viewports.

\clearpage
\balance
\bibliographystyle{ACM-Reference-Format}
\bibliography{main}


\begin{thebibliography}{31}


\ifx \showCODEN    \undefined \def \showCODEN     #1{\unskip}     \fi
\ifx \showDOI      \undefined \def \showDOI       #1{#1}\fi
\ifx \showISBNx    \undefined \def \showISBNx     #1{\unskip}     \fi
\ifx \showISBNxiii \undefined \def \showISBNxiii  #1{\unskip}     \fi
\ifx \showISSN     \undefined \def \showISSN      #1{\unskip}     \fi
\ifx \showLCCN     \undefined \def \showLCCN      #1{\unskip}     \fi
\ifx \shownote     \undefined \def \shownote      #1{#1}          \fi
\ifx \showarticletitle \undefined \def \showarticletitle #1{#1}   \fi
\ifx \showURL      \undefined \def \showURL       {\relax}        \fi
\providecommand\bibfield[2]{#2}
\providecommand\bibinfo[2]{#2}
\providecommand\natexlab[1]{#1}
\providecommand\showeprint[2][]{arXiv:#2}

\bibitem[\protect\citeauthoryear{??}{ocu}{2021}]%
        {oculus}
 \bibinfo{year}{2021 (accessed Feb 26, 2021)}\natexlab{}.
\newblock \bibinfo{title}{Facebook Oculus}.
\newblock \bibinfo{howpublished}{\url{https://www.oculus.com}}.
\newblock


\bibitem[\protect\citeauthoryear{??}{fb}{2021}]%
        {fb}
 \bibinfo{year}{2021 (accessed Feb 26, 2021)}\natexlab{}.
\newblock \bibinfo{title}{Facebook360}.
\newblock \bibinfo{howpublished}{\url{https://facebook360.fb.com/}}.
\newblock


\bibitem[\protect\citeauthoryear{??}{hol}{2021}]%
        {hololens}
 \bibinfo{year}{2021 (accessed Feb 26, 2021)}\natexlab{}.
\newblock \bibinfo{title}{Microsoft Hololens}.
\newblock
  \bibinfo{howpublished}{\url{https://www.microsoft.com/en-us/hololens}}.
\newblock


\bibitem[\protect\citeauthoryear{??}{yt}{2021}]%
        {yt}
 \bibinfo{year}{2021 (accessed Feb 26, 2021)}\natexlab{}.
\newblock \bibinfo{title}{YouTubeVR}.
\newblock \bibinfo{howpublished}{\url{https://vr.youtube.com/}}.
\newblock


\bibitem[\protect\citeauthoryear{Bao, Wu, Zhang, Ramli, and Liu}{Bao
  et~al\mbox{.}}{2016}]%
        {bao2016shooting}
\bibfield{author}{\bibinfo{person}{Yanan Bao}, \bibinfo{person}{Huasen Wu},
  \bibinfo{person}{Tianxiao Zhang}, \bibinfo{person}{Albara~Ah Ramli}, {and}
  \bibinfo{person}{Xin Liu}.} \bibinfo{year}{2016}\natexlab{}.
\newblock \showarticletitle{Shooting a moving target: Motion-prediction-based
  transmission for 360-degree videos}. In \bibinfo{booktitle}{\emph{2016 IEEE
  International Conference on Big Data (Big Data)}}. IEEE,
  \bibinfo{pages}{1161--1170}.
\newblock


\bibitem[\protect\citeauthoryear{Bhise}{Bhise}{2011}]%
        {bhise2011ergonomics}
\bibfield{author}{\bibinfo{person}{Vivek~D Bhise}.}
  \bibinfo{year}{2011}\natexlab{}.
\newblock \bibinfo{booktitle}{\emph{Ergonomics in the automotive design
  process}}.
\newblock \bibinfo{publisher}{CRC Press}.
\newblock


\bibitem[\protect\citeauthoryear{Carlsson and Eager}{Carlsson and
  Eager}{2020}]%
        {carlsson2020had}
\bibfield{author}{\bibinfo{person}{Niklas Carlsson} {and}
  \bibinfo{person}{Derek Eager}.} \bibinfo{year}{2020}\natexlab{}.
\newblock \showarticletitle{Had You Looked Where I'm Looking? Cross-user
  Similarities in Viewing Behavior for 360-degree Video and Caching
  Implications}. In \bibinfo{booktitle}{\emph{Proceedings of the ACM/SPEC
  International Conference on Performance Engineering}}.
  \bibinfo{pages}{130--137}.
\newblock


\bibitem[\protect\citeauthoryear{Chaturvedi, Bijarbooneh, Braud, and
  Hui}{Chaturvedi et~al\mbox{.}}{2019}]%
        {chaturvedi2019peripheral}
\bibfield{author}{\bibinfo{person}{Isha Chaturvedi},
  \bibinfo{person}{Farshid~Hassani Bijarbooneh}, \bibinfo{person}{Tristan
  Braud}, {and} \bibinfo{person}{Pan Hui}.} \bibinfo{year}{2019}\natexlab{}.
\newblock \showarticletitle{Peripheral vision: a new killer app for smart
  glasses}. In \bibinfo{booktitle}{\emph{Proceedings of the 24th International
  Conference on Intelligent User Interfaces}}. \bibinfo{pages}{625--636}.
\newblock


\bibitem[\protect\citeauthoryear{Corbillon, De~Simone, and Simon}{Corbillon
  et~al\mbox{.}}{2017}]%
        {corbillon2017360}
\bibfield{author}{\bibinfo{person}{Xavier Corbillon},
  \bibinfo{person}{Francesca De~Simone}, {and} \bibinfo{person}{Gwendal
  Simon}.} \bibinfo{year}{2017}\natexlab{}.
\newblock \showarticletitle{360-degree video head movement dataset}. In
  \bibinfo{booktitle}{\emph{Proceedings of the 8th ACM on Multimedia Systems
  Conference}}. \bibinfo{pages}{199--204}.
\newblock


\bibitem[\protect\citeauthoryear{He, Qureshi, Qiu, Li, Li, and Han}{He
  et~al\mbox{.}}{2018}]%
        {he2018rubiks}
\bibfield{author}{\bibinfo{person}{Jian He}, \bibinfo{person}{Mubashir~Adnan
  Qureshi}, \bibinfo{person}{Lili Qiu}, \bibinfo{person}{Jin Li},
  \bibinfo{person}{Feng Li}, {and} \bibinfo{person}{Lei Han}.}
  \bibinfo{year}{2018}\natexlab{}.
\newblock \showarticletitle{Rubiks: Practical 360-degree streaming for
  smartphones}. In \bibinfo{booktitle}{\emph{Proceedings of the 16th Annual
  International Conference on Mobile Systems, Applications, and Services}}.
  \bibinfo{pages}{482--494}.
\newblock


\bibitem[\protect\citeauthoryear{Hooft, Vega, Petrangeli, Wauters, and
  Turck}{Hooft et~al\mbox{.}}{2019}]%
        {hooft2019tile}
\bibfield{author}{\bibinfo{person}{Jeroen Van~der Hooft},
  \bibinfo{person}{Maria~Torres Vega}, \bibinfo{person}{Stefano Petrangeli},
  \bibinfo{person}{Tim Wauters}, {and} \bibinfo{person}{Filip~De Turck}.}
  \bibinfo{year}{2019}\natexlab{}.
\newblock \showarticletitle{Tile-based adaptive streaming for virtual reality
  video}.
\newblock \bibinfo{journal}{\emph{ACM Transactions on Multimedia Computing,
  Communications, and Applications (TOMM)}} \bibinfo{volume}{15},
  \bibinfo{number}{4} (\bibinfo{year}{2019}), \bibinfo{pages}{1--24}.
\newblock


\bibitem[\protect\citeauthoryear{Li, Wen, Li, Zhao, Guo, and Wen}{Li
  et~al\mbox{.}}{2016}]%
        {li2016novel}
\bibfield{author}{\bibinfo{person}{Jisheng Li}, \bibinfo{person}{Ziyu Wen},
  \bibinfo{person}{Sihan Li}, \bibinfo{person}{Yikai Zhao},
  \bibinfo{person}{Bichuan Guo}, {and} \bibinfo{person}{Jiangtao Wen}.}
  \bibinfo{year}{2016}\natexlab{}.
\newblock \showarticletitle{Novel tile segmentation scheme for omnidirectional
  video}. In \bibinfo{booktitle}{\emph{2016 IEEE International Conference on
  Image Processing (ICIP)}}. IEEE, \bibinfo{pages}{370--374}.
\newblock


\bibitem[\protect\citeauthoryear{Liu, Zhong, Zhang, Liu, Zhang, Zhang, and
  Gruteser}{Liu et~al\mbox{.}}{2018}]%
        {liu2018cutting}
\bibfield{author}{\bibinfo{person}{Luyang Liu}, \bibinfo{person}{Ruiguang
  Zhong}, \bibinfo{person}{Wuyang Zhang}, \bibinfo{person}{Yunxin Liu},
  \bibinfo{person}{Jiansong Zhang}, \bibinfo{person}{Lintao Zhang}, {and}
  \bibinfo{person}{Marco Gruteser}.} \bibinfo{year}{2018}\natexlab{}.
\newblock \showarticletitle{Cutting the cord: Designing a high-quality
  untethered vr system with low latency remote rendering}. In
  \bibinfo{booktitle}{\emph{Proceedings of the 16th Annual International
  Conference on Mobile Systems, Applications, and Services}}.
  \bibinfo{pages}{68--80}.
\newblock


\bibitem[\protect\citeauthoryear{Lo, Fan, Lee, Huang, Chen, and Hsu}{Lo
  et~al\mbox{.}}{2017}]%
        {lo2017360}
\bibfield{author}{\bibinfo{person}{Wen-Chih Lo}, \bibinfo{person}{Ching-Ling
  Fan}, \bibinfo{person}{Jean Lee}, \bibinfo{person}{Chun-Ying Huang},
  \bibinfo{person}{Kuan-Ta Chen}, {and} \bibinfo{person}{Cheng-Hsin Hsu}.}
  \bibinfo{year}{2017}\natexlab{}.
\newblock \showarticletitle{360 video viewing dataset in head-mounted virtual
  reality}. In \bibinfo{booktitle}{\emph{Proceedings of the 8th ACM on
  Multimedia Systems Conference}}. \bibinfo{pages}{211--216}.
\newblock


\bibitem[\protect\citeauthoryear{Monroy, Lutz, Chalasani, and Smolic}{Monroy
  et~al\mbox{.}}{2018}]%
        {monroy2018salnet360}
\bibfield{author}{\bibinfo{person}{Rafael Monroy}, \bibinfo{person}{Sebastian
  Lutz}, \bibinfo{person}{Tejo Chalasani}, {and} \bibinfo{person}{Aljosa
  Smolic}.} \bibinfo{year}{2018}\natexlab{}.
\newblock \showarticletitle{Salnet360: Saliency maps for omni-directional
  images with cnn}.
\newblock \bibinfo{journal}{\emph{Signal Processing: Image Communication}}
  \bibinfo{volume}{69} (\bibinfo{year}{2018}), \bibinfo{pages}{26--34}.
\newblock


\bibitem[\protect\citeauthoryear{Nasrabadi, Samiei, Mahzari, McMahan, Prakash,
  Farias, and Carvalho}{Nasrabadi et~al\mbox{.}}{2019}]%
        {nasrabadi2019taxonomy}
\bibfield{author}{\bibinfo{person}{Afshin~Taghavi Nasrabadi},
  \bibinfo{person}{Aliehsan Samiei}, \bibinfo{person}{Anahita Mahzari},
  \bibinfo{person}{Ryan~P McMahan}, \bibinfo{person}{Ravi Prakash},
  \bibinfo{person}{Myl{\`e}ne~CQ Farias}, {and} \bibinfo{person}{Marcelo~M
  Carvalho}.} \bibinfo{year}{2019}\natexlab{}.
\newblock \showarticletitle{A taxonomy and dataset for 360° videos}. In
  \bibinfo{booktitle}{\emph{Proceedings of the 10th ACM Multimedia Systems
  Conference}}. \bibinfo{pages}{273--278}.
\newblock


\bibitem[\protect\citeauthoryear{Nguyen, Yan, and Nahrstedt}{Nguyen
  et~al\mbox{.}}{2018}]%
        {nguyen2018your}
\bibfield{author}{\bibinfo{person}{Anh Nguyen}, \bibinfo{person}{Zhisheng Yan},
  {and} \bibinfo{person}{Klara Nahrstedt}.} \bibinfo{year}{2018}\natexlab{}.
\newblock \showarticletitle{Your attention is unique: Detecting 360-degree
  video saliency in head-mounted display for head movement prediction}. In
  \bibinfo{booktitle}{\emph{Proceedings of the 26th ACM international
  conference on Multimedia}}. \bibinfo{pages}{1190--1198}.
\newblock


\bibitem[\protect\citeauthoryear{Ozcinar, Cabrera, and Smolic}{Ozcinar
  et~al\mbox{.}}{2019}]%
        {ozcinar2019visual}
\bibfield{author}{\bibinfo{person}{Cagri Ozcinar}, \bibinfo{person}{Julian
  Cabrera}, {and} \bibinfo{person}{Aljosa Smolic}.}
  \bibinfo{year}{2019}\natexlab{}.
\newblock \showarticletitle{Visual attention-aware omnidirectional video
  streaming using optimal tiles for virtual reality}.
\newblock \bibinfo{journal}{\emph{IEEE Journal on Emerging and Selected Topics
  in Circuits and Systems}} \bibinfo{volume}{9}, \bibinfo{number}{1}
  (\bibinfo{year}{2019}), \bibinfo{pages}{217--230}.
\newblock


\bibitem[\protect\citeauthoryear{Qian, Han, Xiao, and Gopalakrishnan}{Qian
  et~al\mbox{.}}{2018}]%
        {qian2018flare}
\bibfield{author}{\bibinfo{person}{Feng Qian}, \bibinfo{person}{Bo Han},
  \bibinfo{person}{Qingyang Xiao}, {and} \bibinfo{person}{Vijay
  Gopalakrishnan}.} \bibinfo{year}{2018}\natexlab{}.
\newblock \showarticletitle{Flare: Practical viewport-adaptive 360-degree video
  streaming for mobile devices}. In \bibinfo{booktitle}{\emph{Proceedings of
  the 24th Annual International Conference on Mobile Computing and
  Networking}}. \bibinfo{pages}{99--114}.
\newblock


\bibitem[\protect\citeauthoryear{Schmitt, Bronzino, Ayoubi, Martins, Teixeira,
  and Feamster}{Schmitt et~al\mbox{.}}{2019}]%
        {schmitt2019inferring}
\bibfield{author}{\bibinfo{person}{Paul Schmitt}, \bibinfo{person}{Francesco
  Bronzino}, \bibinfo{person}{Sara Ayoubi}, \bibinfo{person}{Guilherme
  Martins}, \bibinfo{person}{Renata Teixeira}, {and} \bibinfo{person}{Nick
  Feamster}.} \bibinfo{year}{2019}\natexlab{}.
\newblock \showarticletitle{Inferring streaming video quality from encrypted
  traffic: Practical models and deployment experience}.
\newblock \bibinfo{journal}{\emph{arXiv preprint arXiv:1901.05800}}
  (\bibinfo{year}{2019}).
\newblock


\bibitem[\protect\citeauthoryear{Shi, Gupta, and Jana}{Shi
  et~al\mbox{.}}{2019}]%
        {shi2019freedom}
\bibfield{author}{\bibinfo{person}{Shu Shi}, \bibinfo{person}{Varun Gupta},
  {and} \bibinfo{person}{Rittwik Jana}.} \bibinfo{year}{2019}\natexlab{}.
\newblock \showarticletitle{Freedom: Fast recovery enhanced vr delivery over
  mobile networks}. In \bibinfo{booktitle}{\emph{Proceedings of the 17th Annual
  International Conference on Mobile Systems, Applications, and Services}}.
  \bibinfo{pages}{130--141}.
\newblock


\bibitem[\protect\citeauthoryear{Strasburger, Rentschler, and
  J{\"u}ttner}{Strasburger et~al\mbox{.}}{2011}]%
        {strasburger2011peripheral}
\bibfield{author}{\bibinfo{person}{Hans Strasburger}, \bibinfo{person}{Ingo
  Rentschler}, {and} \bibinfo{person}{Martin J{\"u}ttner}.}
  \bibinfo{year}{2011}\natexlab{}.
\newblock \showarticletitle{Peripheral vision and pattern recognition: A
  review}.
\newblock \bibinfo{journal}{\emph{Journal of vision}} \bibinfo{volume}{11},
  \bibinfo{number}{5} (\bibinfo{year}{2011}), \bibinfo{pages}{13--13}.
\newblock


\bibitem[\protect\citeauthoryear{Sun, Lu, and Yu}{Sun et~al\mbox{.}}{2017}]%
        {sun2017weighted}
\bibfield{author}{\bibinfo{person}{Yule Sun}, \bibinfo{person}{Ang Lu}, {and}
  \bibinfo{person}{Lu Yu}.} \bibinfo{year}{2017}\natexlab{}.
\newblock \showarticletitle{Weighted-to-spherically-uniform quality evaluation
  for omnidirectional video}.
\newblock \bibinfo{journal}{\emph{IEEE signal processing letters}}
  \bibinfo{volume}{24}, \bibinfo{number}{9} (\bibinfo{year}{2017}),
  \bibinfo{pages}{1408--1412}.
\newblock


\bibitem[\protect\citeauthoryear{T}{T}{1982}]%
        {ohtuski1982}
\bibfield{author}{\bibinfo{person}{Ohtuski T}.}
  \bibinfo{year}{1982}\natexlab{}.
\newblock \showarticletitle{Minimum Dissection of Rectilinear Regions}.
\newblock \bibinfo{journal}{\emph{Proceedings 1982 International Symposium on
  Circuits And Systems (ISCAS)}} (\bibinfo{year}{1982}),
  \bibinfo{pages}{1210--1213}.
\newblock


\bibitem[\protect\citeauthoryear{Wu, Tan, Wang, and Yang}{Wu
  et~al\mbox{.}}{2017}]%
        {wu2017dataset}
\bibfield{author}{\bibinfo{person}{Chenglei Wu}, \bibinfo{person}{Zhihao Tan},
  \bibinfo{person}{Zhi Wang}, {and} \bibinfo{person}{Shiqiang Yang}.}
  \bibinfo{year}{2017}\natexlab{}.
\newblock \showarticletitle{A dataset for exploring user behaviors in VR
  spherical video streaming}. In \bibinfo{booktitle}{\emph{Proceedings of the
  8th ACM on Multimedia Systems Conference}}. \bibinfo{pages}{193--198}.
\newblock


\bibitem[\protect\citeauthoryear{Wu and Sahni}{Wu and Sahni}{1994}]%
        {wu1994fast}
\bibfield{author}{\bibinfo{person}{San-Yuan Wu} {and} \bibinfo{person}{Sartaj
  Sahni}.} \bibinfo{year}{1994}\natexlab{}.
\newblock \showarticletitle{Fast algorithms to partition simple rectilinear
  polygons}.
\newblock \bibinfo{journal}{\emph{VLSI Design}} \bibinfo{volume}{1},
  \bibinfo{number}{3} (\bibinfo{year}{1994}), \bibinfo{pages}{193--215}.
\newblock


\bibitem[\protect\citeauthoryear{Xiao, Zhou, Liu, and Chen}{Xiao
  et~al\mbox{.}}{2017}]%
        {xiao2017optile}
\bibfield{author}{\bibinfo{person}{Mengbai Xiao}, \bibinfo{person}{Chao Zhou},
  \bibinfo{person}{Yao Liu}, {and} \bibinfo{person}{Songqing Chen}.}
  \bibinfo{year}{2017}\natexlab{}.
\newblock \showarticletitle{Optile: Toward optimal tiling in 360-degree video
  streaming}. In \bibinfo{booktitle}{\emph{Proceedings of the 25th ACM
  international conference on Multimedia}}. \bibinfo{pages}{708--716}.
\newblock


\bibitem[\protect\citeauthoryear{Xie, Xu, Ban, Zhang, and Guo}{Xie
  et~al\mbox{.}}{2017}]%
        {xie2017360probdash}
\bibfield{author}{\bibinfo{person}{Lan Xie}, \bibinfo{person}{Zhimin Xu},
  \bibinfo{person}{Yixuan Ban}, \bibinfo{person}{Xinggong Zhang}, {and}
  \bibinfo{person}{Zongming Guo}.} \bibinfo{year}{2017}\natexlab{}.
\newblock \showarticletitle{360probdash: Improving qoe of 360 video streaming
  using tile-based http adaptive streaming}. In
  \bibinfo{booktitle}{\emph{Proceedings of the 25th ACM international
  conference on Multimedia}}. \bibinfo{pages}{315--323}.
\newblock


\bibitem[\protect\citeauthoryear{Xie and Zhang}{Xie and Zhang}{2017}]%
        {xie2017poi360}
\bibfield{author}{\bibinfo{person}{Xiufeng Xie} {and} \bibinfo{person}{Xinyu
  Zhang}.} \bibinfo{year}{2017}\natexlab{}.
\newblock \showarticletitle{Poi360: Panoramic mobile video telephony over lte
  cellular networks}. In \bibinfo{booktitle}{\emph{Proceedings of the 13th
  International Conference on emerging Networking EXperiments and
  Technologies}}. \bibinfo{pages}{336--349}.
\newblock


\bibitem[\protect\citeauthoryear{Yu, Lakshman, and Girod}{Yu
  et~al\mbox{.}}{2015}]%
        {yu2015content}
\bibfield{author}{\bibinfo{person}{Matt Yu}, \bibinfo{person}{Haricharan
  Lakshman}, {and} \bibinfo{person}{Bernd Girod}.}
  \bibinfo{year}{2015}\natexlab{}.
\newblock \showarticletitle{Content adaptive representations of omnidirectional
  videos for cinematic virtual reality}. In
  \bibinfo{booktitle}{\emph{Proceedings of the 3rd International Workshop on
  Immersive Media Experiences}}. \bibinfo{pages}{1--6}.
\newblock


\bibitem[\protect\citeauthoryear{Zhou, Xiao, and Liu}{Zhou
  et~al\mbox{.}}{2018}]%
        {zhou2018clustile}
\bibfield{author}{\bibinfo{person}{Chao Zhou}, \bibinfo{person}{Mengbai Xiao},
  {and} \bibinfo{person}{Yao Liu}.} \bibinfo{year}{2018}\natexlab{}.
\newblock \showarticletitle{Clustile: Toward minimizing bandwidth in 360-degree
  video streaming}. In \bibinfo{booktitle}{\emph{IEEE INFOCOM 2018-IEEE
  Conference on Computer Communications}}. IEEE, \bibinfo{pages}{962--970}.
\newblock


\end{thebibliography}



\end{document}